%% file: main.tex
\documentclass[review,3p,times]{elsarticle}

\usepackage{amssymb}
\usepackage{amsmath}
\usepackage{algpseudocode}
\usepackage{algorithm}
\usepackage{lineno}
\usepackage{bm} 
\usepackage{longtable}
\usepackage{caption}
\usepackage{algorithmicx}

\usepackage{adjustbox}
\usepackage{longtable}
\usepackage{soul}
\usepackage{makecell}
\usepackage{color}
\usepackage{xtab}

\usepackage{float}
\usepackage{nomencl}
\makenomenclature

\usepackage{array}
\newcolumntype{L}{>{\raggedright\arraybackslash}c}

\renewcommand{\hl}[1]{#1} 
\newcommand{\ddk}[2]{\frac{\partial {#1}}{\partial{#2}}}
\newcommand{\dsqdk}[2]{\frac{\partial^2 {#1}}{\partial{#2}\,^2}}
\newcommand{\dkdk}[3]{\frac{\partial^2 {#1}}{\partial{#2}\partial{#3}}}

\journal{Nuclear Engineering and Technology}

\begin{document}

\begin{frontmatter}



\title{Observations Regarding the Construction of Multipoint Kinetics Models from Observational Data}


\author[MITN]{K.G. Howey}

\author[CEA]{Giorgio Valocchi}

\author[MITN]{Dean Price}
\ead{dnprice@mit.edu}

\affiliation[MITN]{organization={Department of Nuclear Science and Engineering, Massachusetts Institute of Technology},
            addressline={60 Vassar St.}, 
            city={Cambridge},
            postcode={02139}, 
            state={MA},
            country={US}}

\affiliation[CEA]{organization={CEA, DES, IRESNE, DER, SPRC },
            addressline={Cadarache}, 
            city={Saint-Paul-lez-Durance},
            postcode={F-13108}, 
            country={France}}

\begin{abstract}
 The multipoint kinetics (MPK) equations extend point kinetics models by tracking neutron populations in lumped reactor regions, offering improved spatial resolution at modest computational cost. 
 \hl{The interactions between the lumped regions are described through coupling coefficients set in a matrix $K$. 
 However, determining $K$} from transient data remains an open problem without established experimental procedures. 
 \hl{The research objective of this work is to identify methods and their corresponding challenges of recovering} $K$ \hl{from observable forward flux measurements during reactor transients.} 
 \hl{Specifically, some novelty lies in} analytical solutions to the Kobayashi MPK formulation that are derived in the case of no precursors and with a single precursor group in spectral form, enabling a mathematically rigorous investigation of eigenmode behavior and parameter observability. 
 Coupling coefficient recoverability is investigated using modal sensitivity analysis and the Hessian condition number of the spectral recovery problem.
 The inclusion of delayed neutron precursors delays this eigenmode divergence but may be insufficient for enabling practical recovery of $K$ in many scenarios. 
 \hl{Most prominently, this work concludes that capturing the behavior of nondominant eigenmodes without oversampling the dominant eigenmode is critical to the recoverability of} $K$ \hl{through transient data.} 
\end{abstract}




\end{frontmatter}


\nomenclature{\(C_m\)}{Delayed neutron precursor concentration in region $m$}
\nomenclature{\(C^\circ_m\)}{Initial delayed neutron precursor concentration in region $m$ ($t=0$)}
\nomenclature{\(E\)}{Error function quantifying the difference between observational data and the model }
\nomenclature{\(H\)}{Loss function Hessian}
\nomenclature{\(K\)}{Matrix of coupling coefficients between lumped regions in the reactor}
\nomenclature{\(k_{mn}\)}{Expected number of neutrons produced at next generation in region $m$ for a single neutron produced at current generation in region $n$}
\nomenclature{\(k^d_{mn}\)}{Expected number of neutrons produced at next generation in region $m$ for a single delayed neutron produced at current generation in region $n$}  
\nomenclature{\(L\)}{Loss function quantifying the misalignment of the observational data and model over a specific temporal interval}
\nomenclature{\(\ell_m\)}{Average time for prompt neutrons anywhere in the reactor to produce neutrons in region $m$}
\nomenclature{\(M^\circ_{b,j}\)}{Overlap of eigenmodes $b$ and $j$ at the initial conditions}
\nomenclature{\(N\)}{Number of lumped regions}
\nomenclature{\(n(t)\)}{Neutron counts}
\nomenclature{\(P_j\)}{$j$-th eigenmode contributions to the fission neutrons source rates }
\nomenclature{\(P^\circ_j\)}{$j$-th eigenmode contributions to the source initial conditions $S^\circ$ }
\nomenclature{\(Q\)}{Eigenvector matrix}
\nomenclature{\(q_j\)}{$j$-th eigenvector}
\nomenclature{\(R_j\)}{$j$-th eigenmode contributions to the delayed neutron precursor concentrations}
\nomenclature{\(R^\circ_j\)}{$j$-th eigenmode contributions to the initial delayed neutron precursor concentrations $C^\circ$} 
\nomenclature{\(S_m\)}{Fission neutron source rate in region $m$ }
\nomenclature{\(S^\circ_m\)}{Initial fission neutron source rate in region $m$ ($t=0$) }
\nomenclature{\(T\)}{Observation interval}
\nomenclature{\(t\)}{Time}
\nomenclature{\(W\)}{Weighting matrix}
\nomenclature{\(\alpha_b\)}{Real $b$-th eigenvalue of $K$}
\nomenclature{\(\hat{\alpha}_b\)}{Model estimated $b$-th eigenvalue of $K$}
\nomenclature{\(\beta_m\)}{Fraction of delayed neutrons produced in region $m$}
\nomenclature{\(\beta_{mn}\)}{Fraction of neutrons produced in the next generation in region $m$ by delayed neutron produced in region $n$}
\nomenclature{\(\kappa\)}{Condition Number}
\nomenclature{\(\lambda\)}{Decay constant }
\nomenclature{\(\rho\)}{Reactivity}
\printnomenclature


\section{Introduction}
\label{sec:intro}
Nuclear reactor dynamics models to support real-time operation can be particularly difficult to obtain given the significant computational cost of some transient simulation methods.
Historically, the point kinetics (PK) equations have provided reasonably accurate characterizations of reactor dynamical behavior without significant computational cost but suffer from a lack of spatial resolution.
To address this shortcoming, the multipoint kinetics (MPK) equations use a similar mathematical approach to the point kinetics equations but allow for tracking neutron populations within lumped core regions.
Unfortunately, this additional resolution introduces multiple quantities that require treatment.
The reactivity, an important quantity in the PK equations, is encoded in a matrix, $K$, whose entries describe the neutron contribution of each lumped region to another.
With the PK equations, reactivity can be measured or calculated with a variety of methods.
For the MPK equations, no clear procedure has been established.
\hl{The novelty of this work lies in building on the concepts presented in} \cite{tommasi2019verification},\hl{ deriving an analytical solution to the MPK equations (albeit idealized), and presenting plausible methods for recovering $K$ from observational data. 
It further indicates some of the challenges and opportunities surrounding the inference of the entries in $K$ from observational data.}

Reactivity for the PK equations can be obtained through various measurement methods \cite{wakabayashi2023introduction}. 
One method type employs experiments beginning with a reactor in a low-power, critical state. 
The positive period method measures small amounts of positive reactivity insertion by withdrawing a control rod, thus making the reactor supercritical, and measuring the reactor period.
A representative use of this method is documented for the SPERT-III E-core in \cite{potenza1966spert}, where a chemical reactivity shim is incorporated to determine total control rod worths.
Similarly, the rod drop method measures large negative reactivity insertions. 
Reactivity can also be determined from a subcritical state with an external neutron source. 
In the neutron source multiplication method, a subcritical reactor can be subjected to an external neutron source and subcritical multiplication can be used to determine the reactivity characteristics.
The source jerk method also begins in a subcritical steady state with an external source, and parallels the rod drop method. 
More recently, this method has been used to perform subcritical reactivity measurements of the Jordan subcritical assembly \cite{jarrah2018experimental}.
Finally, given the neutron counts over a transient ($n(t)$), the inverse kinetics method utilizes the PK equations to back calculate the reactivity over a transient ($\rho (t)$) employing numerical methods such as Euler or Runge-Kutta \cite{monteiro2024differential, ferney2024benchmarking}.
This range of measurement methods has helped support the reliable use of the point kinetics equations.
Reactivity can often be validated experimentally, and in some cases it can be obtained directly from measurement in the absence of computational modeling.
The multipoint kinetics equations do not benefit from the same level of experimental support.

In current multipoint kinetic theory models, a reactor is divided into $N$  lumped regions, forming a system of first order differential equations that includes the neutron sources and delayed neutron precursors of each region. 
An early formalization of multipoint kinetic theory was developed in 1958 by R. Avery \cite{avery1958theory}. 
Avery’s model uses a combination of  fission and transport operators, forward fluxes, and adjoint fluxes, each applied over specific and global regions to quantify the neutrons from each region $m$ contributing to each region $n$, requiring $N^2$ equations to characterize the neutron populations over time. 
An alternative popular modeling approach is presented in 1992 by Kobayashi \cite{kobayashi1992rigorous}, where Avery’s application of the adjoint function as a weighting function is replaced with a Green’s function.
One important distinction is that Kobayashi's model reduced the neutron population equations from $N^2$ to $N$ equations. 
A third major MPK approach was developed by Shimjith et al. in 2010 \cite{shimjith2010space}. 
Shimjith’s model is derived from the diffusion approximation and defines its coupling coefficients through a ratio of geometric and material properties of the reactor. 
Recent discussions in multipoint kinetics include the impact of region heterogeneity and expansion to multiple energy groups, as in Dulla et al. \cite{Dulla2017Multipoint}. 
Valocchi et al. \cite{valocchi2020reduced} applied Avery and Kobayashi MPK models to step-change transients in coupled fast--thermal and sodium-cooled fast-reactor systems, showing that MPK can represent rapid shape redistribution impacting the magnitude of conventional prompt-jump behavior.
\hl{Additionally, Hui} \cite{Hui03082025} \hl{applies a two-region model and incorporates extended state observer integration schemes to accurately simulate a PWR system during load-following operation.} 
\hl{PWR load following was also investigated by Zaidabadi nejad and Ansarifar through simulating operation using an adaptive controller with a Lyapunov approach} \cite{ansarifar2017adaptive}. 
\hl{Related work applied adaptive Kalman filtering to estimate axially distributed reactor states from available measurements using a nodal core model} \cite{HUI2025128006}. 
In this note, the discussion is presented in the context of the Kobayashi MPK equations, but concepts may be applied to alternative formulations.

\hl{The objective of this work is to present methods and their respective challenges surrounding obtaining coupling coefficients for the multipoint kinetics equations from observational transient data alone.} 
To limit the scope of the current work, ``observational data'' is used to refer to the potentially measurable forward flux solution.
The main observation from this note is that quantities associated with the higher-order modes of the spatial flux distribution may be difficult to obtain experimentally due to the rapid decay of these modes.
This discussion is provided to inform the development of experimental procedures for determining parameters for the MPK equations.
This note may also have value in that it presents analytical solutions to the MPK equations under simplifying assumptions that permit a particular spectral treatment of the dynamics.

\section{Equations and Models}
\label{sec:method}

\subsection{Kobayashi Multipoint Kinetic Equations}
Using the approach provided in \cite{kobayashi1992rigorous}, the MPK equations with one precursor group applied to $N$ lumped regions can be stated as 
\begin{gather} 
\ell_m \dfrac{d S_m}{d t} = \displaystyle\sum_{n=1}^{N} (1 - \beta_{mn}) k_{mn} S_n(t) - S_m(t) + \lambda \sum_{n=1}^{N} k^d_{mn} C_n(t), \label{EQ:MPK-dS} \\
\frac{dC_m}{dt} = 
\beta_m S_m(t) - 
\lambda C_m(t), \quad  \text{for } m  = 1, ..., N
\label{EQ:MPK-dC}
\end{gather}
to form a coupled system of $2N$ equations.
Here, $S_m(t)$ is the fission neutron source rate in region $m$, $C_m(t)$ is the delayed neutron precursor concentration in region $m$, and $\lambda$ is the representative decay constant of the single precursor group. 
Initial conditions of each source and precursor in region $m$ are given as $S_m^\circ$ and $C_m^\circ$, respectively. 
Most commonly, the precursors are divided into six or eight groups, as seen in Keepin \cite{keepin1965physics} and Rudstam et al. \cite{rudstam2002delayed}, respectively; however, to analytically investigate kinetic behavior in more depth, a single precursor group is assumed. 
Then, $\beta_{mn}$ is the fraction of neutrons produced in the next generation in region $m$ by delayed neutrons produced in the current generation in region $n$, and $\beta_m$ is the raw delayed neutron fraction in region $m$.
The prompt generation time, $\ell_m$, is the average time for prompt neutrons anywhere in the reactor to reach region $m$ and produce next generation neutrons. 
The coupling coefficient $k_{mn}$ is the expected number of neutrons produced at next generation in region $m$ for a single neutron produced at current generation in region $n$.
\hl{The coupling coefficient for delayed neutrons $k^d_{mn}$ is the expected number of neutrons produced at next generation in region $m$ for a single delayed neutron produced at current generation in region $n$.} 

 
To frame the discussion provided in the rest of the paper, assume that the reactor regions are geometrically and materially similar, such that $\beta_{mn} = \beta$ and $\ell_m = \ell$, thus removing effects caused by asymmetries. 
Differences between $k_{mn}$ and $k^d_{mn}$ may arise from the different spectral distribution of prompt and delayed fission neutrons.
However, for simplicity, assume $k_{mn} = k^d_{mn}$ for this note.
The Kobayashi equations can then be simplified by setting the equations in matrix form, as 
\begin{gather} 
\dfrac{d S}{d t} = \Big(\big((1-\beta) K -I\big)S(t) +  \lambda K C(t) \Big)\frac{1}{\ell}
 \label{EQ:MPK-dS2} \\
\dfrac{d C}{d t} = \beta S(t) - \lambda C(t)
 \label{EQ:MPK-dC2}
\end{gather}
with initial conditions $S^\circ\in\mathbb{R}^N$ and $C^\circ\in\mathbb{R}^N$ and where $K\in \mathbb{R}^{N\times N}$ is a matrix of the coupling coefficients, in which the coefficient $k_{mn}$ is in row $m$, column $n$, and $I\in \mathbb{R}^{N\times N}$ is the identity matrix. 
Additionally, $S\in \mathbb{R}^N$ and $C\in \mathbb{R}^N$ are now vector-valued quantities containing the source rate and precursor concentrations within each region, respectively.
For a further simplified case to more directly observe the coupling behavior between regions, a form of the MPK equations with no delayed neutron precursors can be presented,
\begin{gather} 
\dfrac{d S}{d t} = \left(\frac{K-I}{\ell}\right)S(t),
\label{Eq:MPK-dprompt}
\end{gather}
with initial conditions $S^\circ\in\mathbb{R}^N$. This simplification will be used as a platform to temporarily discuss some concepts that will then be applied to the precursor-present case given in Equations \ref{EQ:MPK-dS2} and \ref{EQ:MPK-dC2}.
\subsection{Model Discovery from Observational Data}
To recover model parameters such as the coupling coefficients from observational data, the match of the model to the data must be quantified, in this work calculated through a loss function. 
The notation $\hat{S}(t; \hat{\alpha})$ is used to describe the modeled source as a function of time $t$, parameterized by model  parameters $\hat{\alpha}$. 
The difference with some observational data, $S(t; \alpha)$, formed with hidden true model parameters, $\alpha$,  is notated \hl{with a residual function} as 
\begin{equation}
    E(t;\hat{\alpha}) = S(t;\alpha)-\hat{S}(t;\hat{\alpha}),
    \label{Eq:modelDiff}
\end{equation}
where $E(t;\hat{\alpha})\in \mathbb{R}^{N}$, $S(t;\alpha)\in \mathbb{R}^{N}$ and $\hat{S}(t;\hat{\alpha}) \in \mathbb{R}^{N}$. 
With this, the loss, $L(\hat{\alpha})$, over a specific time-interval, $t \in [0, T]$, can be characterized as
\begin{equation}
    L(\hat{\alpha})=\int_0^T E(t;\hat{\alpha})^T W E(t;\hat{\alpha})dt .
    \label{Eq:Loss}
\end{equation}
To allow for generality, $W\in \mathbb{R}^{N\times N}$ is a diagonal matrix in which the elements correspond to weights of different reactor regions. 
This form of the loss function ensures the differences are summed regardless of sign.


\subsection{Spectral Reformulation of Loss Function}
\label{sec:specre}
Simple approaches for inferring $K$ from observational data may pursue recovering $K$ on an entry-wise basis.
With this entry-wise approach, all individual entries in $K$ will form the set of unknown model parameters, $\hat{\alpha}$, and optimization methods will seek to minimize $L(\hat{\alpha})$ by modifying the entries in $K$.
In this note, an alternative approach is used where $K$ is recovered on a spectral basis.
With this spectral approach, $\hat{\alpha}$ will be composed of the eigenvalues of $K$.
For the purposes of this analysis, it will be assumed that the eigenvectors of $K$ are known quantities such that $K$ can be assembled if the eigenvalues can be recovered.
As will be shown mathematically in Section \ref{sec:cfms}, the components of an eigenvector describe the fractional power relation of the regions, and the eigenvalues describe at what rate each eigenmode exponentially grows or decays, the dominant of which corresponds to the long-time spatial distribution.
Although this does not represent a real-world inference scenario, as the eigenvectors of $K$ are assumed to be known, the discussion in this idealized setting may still yield insights applicable to more general approaches for inferring $K$ from observational data.
For the remainder of this analysis, $K$ is assumed to be diagonalizable over
$\mathbb{R}$, with a complete set of real eigenvectors and corresponding real
eigenvalues.
In practice, real coupling matrices may have complex eigenvalues, as in Table 6 of Tommasi and Palmiotti \cite{tommasi2019verification}. 
In that example, the imaginary parts are sufficiently small that modal decay dominates the associated oscillations, so the observability argument developed here should remain qualitatively valid.

\section{Closed-Form Modal Solutions}
\label{sec:cfms}
\subsection{Precursor-Free Case}
Using an eigendecomposition of $K = Q \,\text{diag}\left(\alpha\right)Q^{-1}$ where ${\alpha} \in \mathbb{R}^N$ is now specified as a vector of eigenvalues of $K$ (as discussed in Section \ref{sec:specre}) and $Q \in \mathbb{R}^{N \times N}$ is the matrix whose columns are the corresponding right eigenvectors, introduce a change of basis 
\begin{equation}
    P(t)=Q^{-1}S(t;\alpha),
    \label{eq:changeofBasis}
\end{equation}
\hl{so that the fission neutron source rates are now described in the eigenbasis.} 
Applying the above change of basis to  Equation \ref{Eq:MPK-dprompt} results in 
\begin{gather}
\label{eq:anahat}
{S}(t; {\alpha}) = \sum_{j=1}^N P_j^\circ e^{\frac{{\alpha}_j - 1}{\ell}t} q_j
\end{gather}
with initial conditions 
\begin{gather}
\begin{bmatrix}
P_1^\circ, \; P_2^\circ, \; \dots \; P_N^\circ
\end{bmatrix}^T \equiv Q^{-1} S^\circ.
\label{eq:initialP}
\end{gather}
where $\alpha_j$ is the $j$-th eigenvalue corresponding to the $j$-th right eigenvector $q_j\in \mathbb{R}^N$, which is the $j$-th column of $Q$. 
The exponential factor describes the decay or growth of the $j$-th eigenmode over time.
\hl{The left-hand side of Equation} \ref{eq:initialP} \hl{is a column-vector in $\mathbb{R}^N$ corresponding to the source initial conditions restated in the new eigenbasis.
$P_j^\circ$ is the $j$-th element of that vector and a constant that corresponds to $j$-th eigenmode contributions to the initial conditions $S^\circ$. } 
Thus the summation of $P_j^\circ q_j$ over $N$ describes the initial conditions $S^\circ$ in the eigenbasis. 
In this form, it is clear that the behavior of an eigenmode is independent of the other eigenmodes, the elements of which may be summed to obtain regional solutions.
Using Equation \ref{Eq:modelDiff} and Equation \ref{Eq:Loss}, the loss function in the case of no precursors becomes
\begin{align}
L(\hat{\alpha})
&= \int_0^T \left(\sum_{h=1}^N P_h^\circ \left(e^{\frac{{\alpha}_h - 1}{\ell}t} - e^{\frac{\hat{\alpha}_h - 1}{\ell}t}\right)q_h^T W \sum_{j=1}^N P_j^\circ \left(e^{\frac{{\alpha}_j - 1}{\ell}t} - e^{\frac{\hat{\alpha}_j - 1}{\ell}t}\right) q_j \right) dt
\label{Eq:LossNoC}.
\end{align}
\hl{The above described derivation process is summarized in a flowchart in} \ref{app:flo-loss}. 
The gradient can then be found to be
\begin{align}
\frac{\partial}{\partial \hat{\alpha}_b}  L(\hat{\alpha})
  =
-\frac{2}{\ell}\sum_{j=1}^N M_{b,j}^\circ
\left(I_1\left(\frac{\hat{\alpha}_b + \alpha_j - 2}{\ell}, T \right) - I_1\left(\frac{\hat{\alpha}_b + \hat{\alpha}_j - 2}{\ell}, T \right)\right),
\end{align}
where
\begin{gather}
M_{b,j}^\circ \equiv P_b^\circ P_j^\circ q_b^T W q_j
\end{gather}
captures the overlap of the eigenmodes at the initiation of the transient and 
\begin{gather}
I_n(\gamma, T) \equiv \int_0^T t^n e^{\gamma t} dt.
\label{Eq:In}
\end{gather}
Integrals of this type have analytical expressions provided in Section 2.32 of \cite{gradshteyn2014table}.
Derivation of the gradient and following Hessian can be found in \ref{app:noC}.
The Hessian can be expressed as 
\begin{gather}
\frac{\partial^2}{\partial \hat{\alpha}_b \partial \hat{\alpha}_c}  L(\hat{\alpha})
=
-\frac{2}{\ell^2}\delta_{b,c}\sum_{j=1}^N M_{b,j}^\circ
\left(
I_2\left(\frac{\hat{\alpha}_b+\alpha_j-2}{\ell}, T\right)
-
I_2\left(\frac{\hat{\alpha}_b+\hat{\alpha}_j-2}{\ell}, T\right)
\right)
\label{Eq:HessNoC}\\
+
\frac{2}{\ell^2}M_{b,c}^\circ
I_2\left(\frac{\hat{\alpha}_b+\hat{\alpha}_c-2}{\ell}, T\right),
\notag
\end{gather}
\hl{where $\delta_{b,c}$ is the Kronecker delta function.} 

\subsection{Precursor-Present Case}
Now investigate the case in which the delayed neutrons are characterized by a single precursor group. 
To begin, introduce $R(t)=Q^{-1}C(t;\alpha)$, similar to Equation \ref{eq:changeofBasis}. 
Implementing this change of basis into Equation \ref{EQ:MPK-dS2} and Equation \ref{EQ:MPK-dC2} results in $N$ pairs of coupled equations. The $j$-th coupled equations are
\begin{gather}
\ell \frac{d P_j}{dt} = \left(\left( 1 - \beta\right)\alpha_j - 1\right) P_j(t) +  \lambda \alpha_j R_j(t) \\
\frac{dR_j}{dt} = \beta P_j(t) - \lambda R_j(t),
\end{gather}
where $P_j$ and $R_j$ correspond to the $j$-th eigenmode and are the $j$-th components of $P(t)$ and $R(t)$, respectively.
The source $P_j$ and precursors $R_j$ are coupled but are independent from other eigenmodes.
Additionally, for the initial precursor conditions, define
\begin{gather}
\begin{bmatrix}
R_1^\circ, \; R_2^\circ, \; \dots \; R_N^\circ
\end{bmatrix}^T
\equiv Q^{-1}C^\circ
\end{gather}
where 
$R_j^\circ$ is a constant that corresponds to $j$-th eigenmode contributions to the initial conditions $C^\circ$
so that $P_j^\circ$ and $R_j^\circ$ give the initial values of $P_j(t)$ and $R_j(t)$, respectively. 
In form, these equations are comparable to conventional point kinetics equations \cite{duderstadt1975nuclear}.
Therefore, we can employ a characteristic equation method to yield
\begin{gather}
P_j(t) = \xi_j^+ \exp(\omega_j^+ t) - \xi_j^- \exp(\omega_j^- t), \\
R_j(t) = \zeta_j^+ \exp(\omega_j^+ t) - \zeta_j^- \exp(\omega_j^- t).
\end{gather}
where
\begin{gather}
\omega_{j}^{\pm}
=
\frac{1}{2}
\left(
\frac{(1-\beta)\alpha_j-1}{\ell}-\lambda
\pm
\sqrt{
\left(\lambda-\frac{(1-\beta)\alpha_j-1}{\ell}\right)^2
+\frac{4\lambda(\alpha_j-1)}{\ell}
}
\right),
\end{gather}
with $\omega^{\pm}_j$ corresponding to the $j$-th eigenmode and
\begin{align}
    \zeta_j^\pm = \frac{\beta P_j^\circ - \left(\omega_j^\mp + \lambda\right)R_j^\circ}{\omega_j^+ - \omega_j^-}, \hspace{1.5cm}
    \xi_j^\pm = \frac{\omega_j^\pm + \lambda}{\beta}\zeta_j^\pm.
\end{align}
Then, transforming back to the original basis gives
\begin{gather}
S(t;\alpha) =
\sum_{j=1}^N
\left(
\xi_j^+ \exp( \omega_j^+ t)
-
\xi_j^- \exp( \omega_j^- t)
\right)q_j
\\
C(t;\alpha) =
\sum_{j=1}^N
\left(
\zeta_j^+ \exp( \omega_j^+ t)
-
\zeta_j^- \exp( \omega_j^- t)
\right)q_j.
\end{gather}
To derive the loss function for the single precursor case, define $E(t;\hat{\alpha})$ for this case as
\begin{align}
    E(t;\hat{\alpha})  &= \sum_{j=1}^N
    \left[
    \left(\xi_j^+ \exp( \omega_j^+ t)-\xi_j^- \exp( \omega_j^- t)\right)
    - \left(\hat{\xi}_j^+ \exp( \hat{\omega}_j^+ t)-\hat{\xi_j}^- \exp( \hat{\omega}_j^- t)\right)
\right]q_j,
\label{Eq:E-1C}
\end{align}
where $\hat{\xi_j}^\pm$ and $\hat{\omega}_j^\pm$ are coefficients of a similar formula as ${\xi_j}^\pm$ and ${\omega}_j^\pm$ but utilize the approximated model eigenvalue $\hat{\alpha}_j$. 
This leads to the loss function for one delayed neutron precursor group being expressed as
\begin{align}
    L(\hat{\alpha})=
    \int_0^T \sum_{h=1}^N&\left[
    \left(\xi_h^+ \exp( \omega_h^+ t) - \hat{\xi}_h^+ \exp( \hat{\omega}_h^+ t)\right)
    - \left(\xi_h^- \exp( \omega_h^- t) -\hat{\xi}_h^- \exp( \hat{\omega}_h^- t) \right)\right]q_h^T \\
    &W\sum_{j=1}^N\left[
    \left(\xi_j^+ \exp( \omega_j^+ t) - \hat{\xi}_j^+ \exp( \hat{\omega}_j^+ t)\right)
    - \left(\xi_j^- \exp( \omega_j^- t) -\hat{\xi}_j^- \exp( \hat{\omega}_j^- t)\right)
\right]q_j dt.
\end{align}
Expressions and their associated derivations for the gradient and Hessian can be found in \ref{app:1C}.

\section{Demonstration Application}
\label{sec:demo_app}
To investigate eigenmode behavior and the possibility of coupling coefficient recovery, the above methods were applied to the  transient published in \cite{tommasi2019verification}, which was also analyzed using Avery and Kobayashi MPK models in \cite{valocchi2020reduced}.
For this transient, a sodium-cooled fast reactor with $2\pi/3$ rotational symmetry is initially critical with a symmetric power distribution. To initiate the transient, a control rod is lifted 35 cm, inserting a reactivity of $\rho=0.00301$.
The reactor is divided into three similar regions, and the relevant parameters used to obtain the following results can be found in Table \ref{tab:Tommasi_params}.
\hl{Further details about the transient can be found in} \cite{tommasi2019verification}. 
Using the above transient, the behavior in the case of no precursors, one precursor group with $\lambda=1$ s$^{-1}$, and one precursor group with $\lambda=0.01$ s$^{-1}$ present is considered. 
An equal weight for all three regions ($W=I$) are used for the loss-function calculations.

\begin{table}[h]
\centering
\caption{Reactor System Parameters and Transition Matrix}
\label{tab:Tommasi_params}
\begin{tabular}{ll}
\hline
\textbf{Variable} & \textbf{Value / Matrix Elements} \\ 
\hline
Initial State ($S^\circ$) & $[1/3 \quad 1/3 \quad 1/3]$ \\
Mean Neutron Lifetime ($\ell$) & $4.30033333 \times 10^{-7}$ s\\ 
Transition Matrix ($K$) & 
$\begin{bmatrix} 
0.93858224 & 0.04424152 & 0.04337424 \\ 
0.03513004 & 0.91835555 & 0.03786149 \\ 
0.03391385 & 0.03731321 & 0.9186247 
\end{bmatrix}$ \\ 
\hline
\end{tabular}
\end{table}

\section{Results}
\label{sec:results}
\subsection{\hl{Eigenmodal Influence}} 
\begin{figure}[htb!]
\centering
\includegraphics[width=0.7\textwidth]{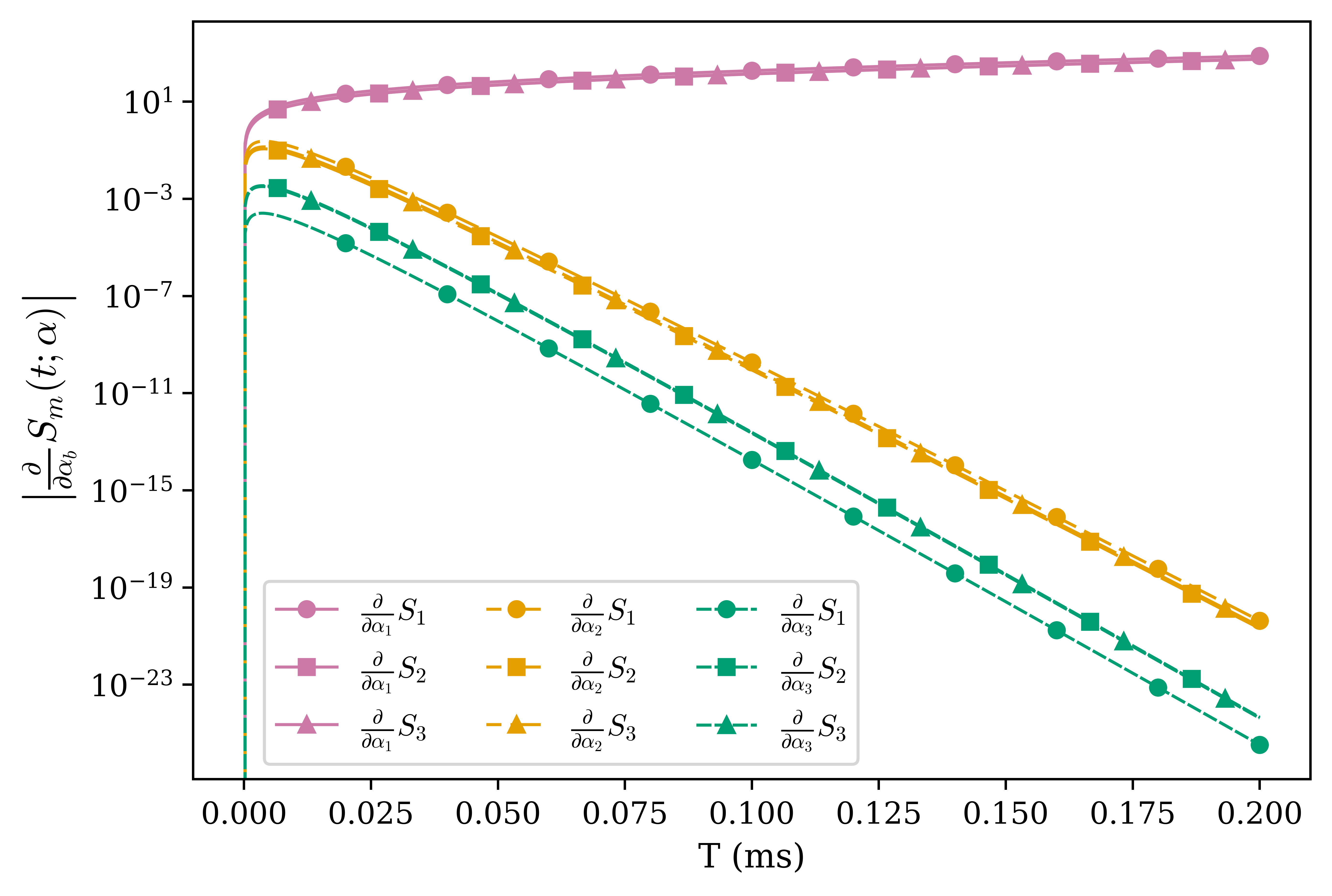}
\caption{Eigenmode sensitivity behavior for the precursor-free MPK model applied to the transient described in Section \ref{sec:demo_app}.}
\label{fig:promptEig}
\end{figure}

Following the spectral reformulation discussed in Section \ref{sec:specre}, it can be readily understood that identifying any information associated with the $j$-th eigenmode relies on the influence of this eigenmode on the observed source solution. 
Therefore, this section will use the sensitivity of the source to each eigenmode (expressed as $\frac{\partial}{\partial \alpha_a} S_m(t;\alpha)$ and derived from Equation \ref{eq:anahat}) to draw some conclusions regarding the observability of the corresponding eigenvalue from transient source measurements.
First, the eigenmode behavior within each region without considering any precursors was explored for the transient described in Section \ref{sec:demo_app}. 
\hl{Figure} \ref{fig:promptEig} \hl{displays the time-dependent sensitivity of the neutron source in each region to each reactor eigenmode.
As a reminder, the subscript on $S$ refers to a specific lumped region and the subscript on $\alpha$ refers to a reactor eigenmode.} 
As the initial conditions are independent of any property of $K$, all sensitivities begin from 0 and rapidly rise within the first few microseconds.
More notably, the influence of the higher order eigenvalues on the sources decays rapidly after their peak.
By 0.1 ms into the transient, the influence of the dominant and least dominant eigenmode on a regional source differs by about 15 orders of magnitude. 
This significant difference in timescales is also reflected in Figure 10 of Valocchi et al. \cite{valocchi2020reduced} for the eigenvalues associated with the full Avery and Kobayashi models.
Recall that the coupling coefficient matrix $K$ can be described in terms of its eigenmodes. 
In this precursor-free case, the rapid decay of the higher-order eigenmodes implies that the corresponding eigenvalues are effectively unrecoverable from sensor data, as their contributions to the observed source solution vanish before they can be resolved experimentally.

\begin{figure}[htb!]
\centering
\includegraphics[width=0.5\textwidth]{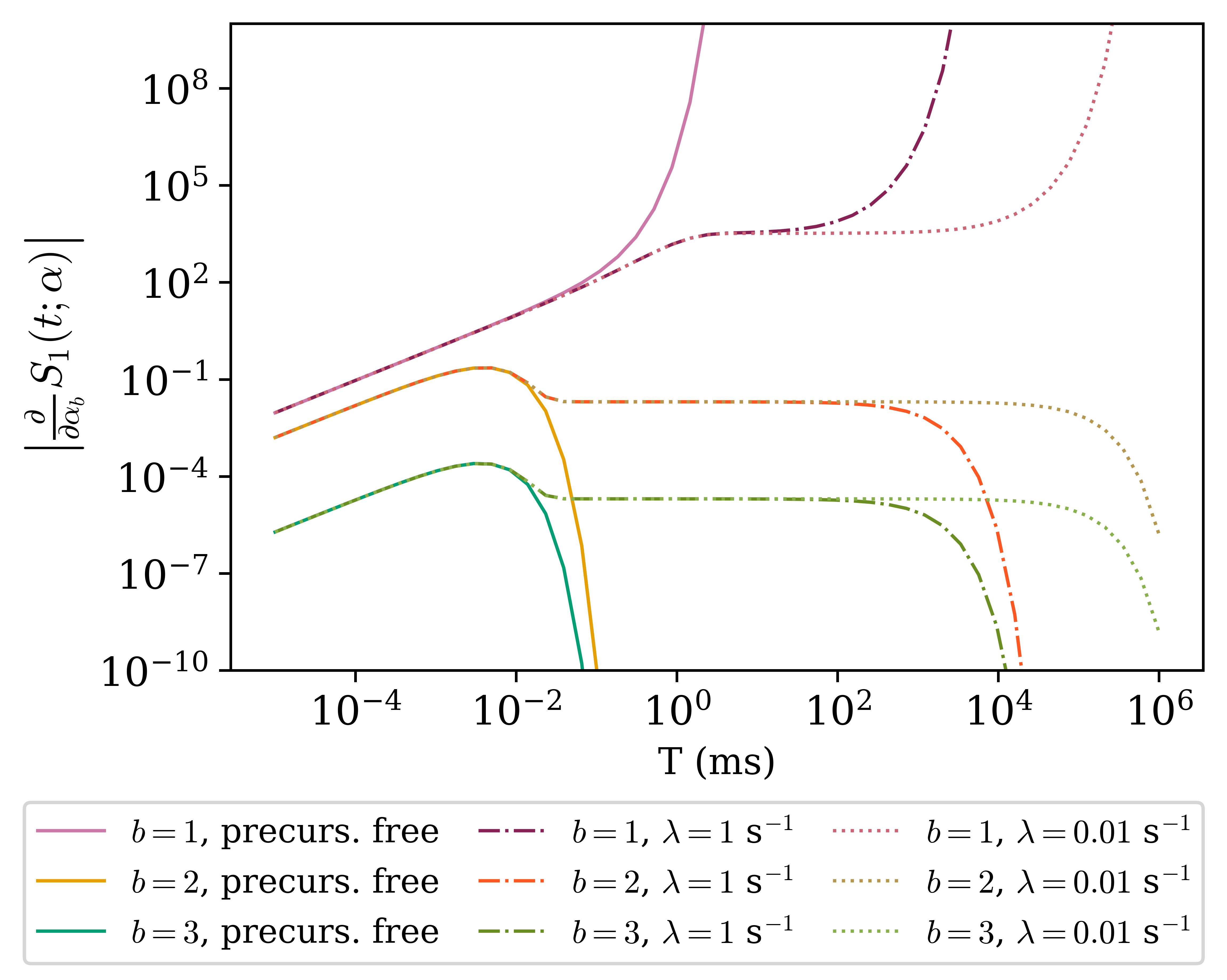}
\caption{Comparison of eigenmode sensitivity behavior for a single region across MPK models with differing precursor group decay constants.}
\label{fig:lambdaseig}
\end{figure}

Of course, real nuclear systems do have delayed neutron precursors.
Figure \ref{fig:lambdaseig} compares the effect of the precursor decay constant \hl{for a single delay group} on  the eigenmode sensitivities in Region 1, where $\beta=367.67$ pcm. 
Here, the temporal axis is given in logarithmic form as more complex behavior occurs across multiple timescales.
Again, the rise in the sensitivity within the first few microseconds can be observed.
However, with the addition of a precursor group, the sensitivities do not immediately decay and remain closer for orders of magnitude longer. 
In the case where $\lambda=1$ s$^{-1}$, at 0.1 ms the dominant and least dominant eigenmodes differ by only seven orders of magnitude, substantially improving the possibility of recovery from observational data. 
In the case where $\lambda=0.01$ s$^{-1}$, the eigenmodes differ similarly at 0.1 ms but also maintain this proximity until 10000 ms or 10 seconds, offering a larger possible observation window.  
This aligns with expectation, as a smaller decay constant indicates a longer precursor half-life, thus the delayed-neutron contribution is more slowly returned to the source and persists for more time.
The similar values of the temporary sensitivity plateau observed in the precursor present cases are due to a similar delayed neutron fraction. 
\hl{Although only a single precursor group is considered here, the qualitative effect of delayed-neutron precursors extending the observability of nondominant eigenmodes is expected to persist in models with multiple precursor groups.}

\subsection{Conditioning of Spectral Recovery Problem}
The condition number \hl{($\kappa$)} of the Hessian of the loss function \hl{($H$)} evaluated at the true spectrum of $K$, 
\begin{gather}
\kappa(\alpha) = \|H^{-1}(\alpha)\|_2 \|H(\alpha)\|_2,
\end{gather}
can provide insight into the convergence or numerical stability for many optimization methods.
For gradient-based methods, $\kappa(\alpha)$ affects the rate of convergence and, as with Newton's method, can amplify the effect of numerical or observational errors \cite{BoydVandenberghe2004}.
In a practical implementation, observational errors would arise from measurement uncertainty.
This conditioning also plays a role in metaheuristic algorithms \cite{ShirYehudayoff2020}.
Generally, it should be understood that $\kappa(\alpha)$ is a method-agnostic measure of the difficulty of the loss function minimization task.

\begin{figure}[h!]
\centering
\includegraphics[width=0.5\textwidth]{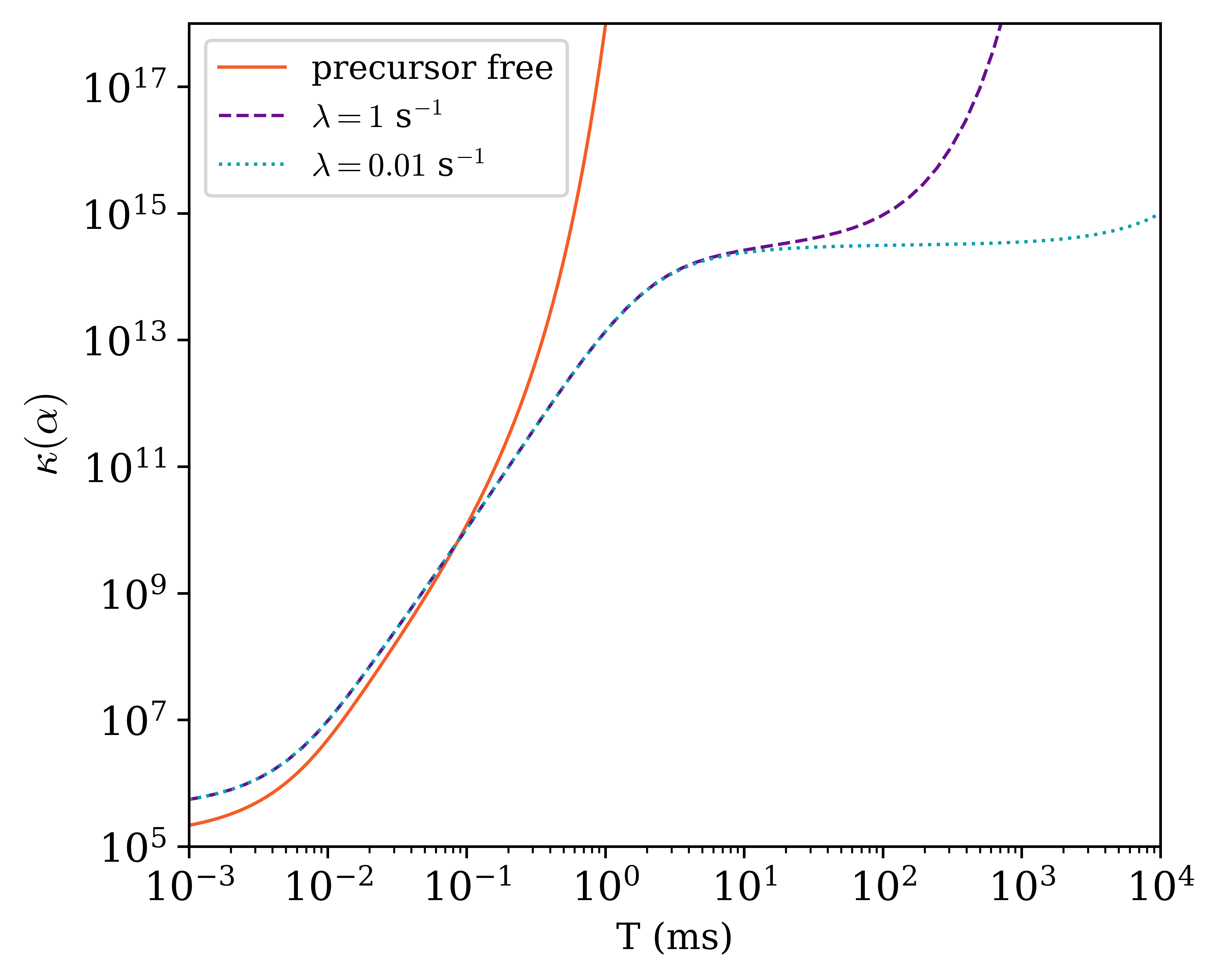}
\caption{Hessian condition number evaluated at the true spectrum, $\kappa(\alpha)$, as a function of observation interval $T$ across MPK models.}\label{fig:CondNum}
\end{figure}

Therefore, Figure \ref{fig:CondNum} reports $\kappa(\alpha)$ as a function of the observation interval $T$ for three precursor configurations.
This allows the difficulty of the spectral recovery problem to be compared across transient observation windows.
The introduction of precursors inhibits the growth of the condition number by orders of magnitude as expected.
However, this inhibition does not occur instantly.
By the time precursor effects substantially limit the growth of $\kappa(\alpha)$, the condition number has already increased by five orders of magnitude from $T\approx10^{-3}$ ms to $T\approx10^{-1}$ ms and the higher-order eigenmodes have already become minimally represented in the source.

It should be emphasized that the observability of $\alpha$ is not an algorithmic challenge, but mainly a measurement challenge.
In the presence of measurement uncertainty, errors in the observational data may be amplified by the conditioning of the loss function, which in turn affects the convergence and stability of the recovery problem.
In real-world applications, this creates a trade-off in the selection of the observation interval $T$.
A small $T$ may preserve information from the higher-order eigenmodes but suffer from significant detector uncertainty, while a large $T$ may reduce detector uncertainty but lead to poor conditioning and loss of modal information.
As a result, $\alpha$ may be unrecoverable from straightforward transient measurements alone.
This challenge was recognized in the MUSE-4 experiments \cite{MUSE_D8_2005}, where repetitive small-timescale transients were performed and the detector data were combined to obtain a functionally small $T$ with reduced detector noise.
Discussion of this experiment within the context of MPK treatments can be found in \cite{Valocchi_2020}.


\subsection{Application of Newton's Method}
\label{subsec:NM}
To illustrate the concepts discussed in the previous section, this section applies Newton's method \cite{nesterov2018lectures}  to the $\alpha$ recovery problem.
If $\hat{\alpha}^{(l)}\in\mathbb{R}^N$ denotes the approximation to $\alpha$ at Newton iteration $l$,
the Newton update was written as
\begin{gather}
\hat{\alpha}^{(l+1)}=\hat{\alpha}^{(l)}- H^{-1}\left(\hat{\alpha}^{(l)}\right) \nabla L\!\left(\hat{\alpha}^{(l)}\right),
\end{gather}
where $H(\hat{\alpha}^{(l)})\in\mathbb{R}^{N\times N}$ is the Hessian of the loss function evaluated at $\hat{\alpha}^{(l)}$ and $\nabla L(\hat{\alpha}^{(l)})\in\mathbb{R}^{N}$ is the corresponding loss-function gradient.
The initial guess $\hat{\alpha}^{(0)}$ was chosen to be unity so that the exponential coefficient $\frac{\hat{\alpha}_j^{(0)}-1}{\ell}=0 \, \forall j$.

\begin{figure}[h!]
\centering
\includegraphics[width=0.5\textwidth]{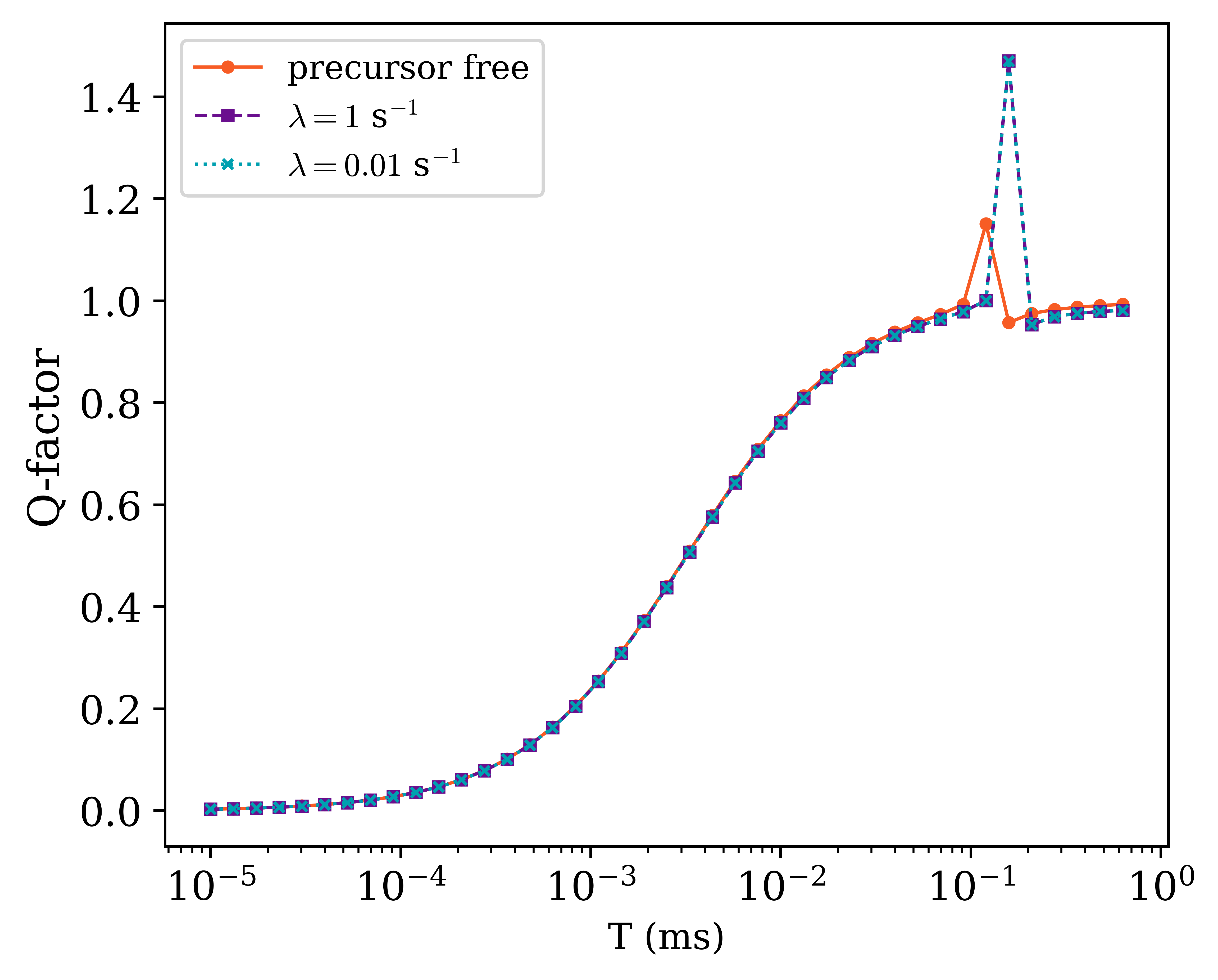}
\caption{Initial Q-factor for Newton's method as a function of observation interval $T$ across MPK models.}\label{fig:firststepratio}
\end{figure}

As an indicator of the convergence potential of Newton's method, the initial Q-factor in the 1-norm was investigated.
An initial Q-factor \cite{burden1985numerical}  can be defined as 
\begin{equation}
\text{Q-factor} = \frac{\|\hat{\alpha}^{(1)}-\alpha\|_1}{\|\hat{\alpha}^{(0)}-\alpha\|_1}.
\end{equation}
This Q-factor is plotted as a function of transient observation interval $T$ in Figure \ref{fig:firststepratio}.
As $T$ increases, the improvement towards $\alpha$ in the first iteration decreases, illustrating how the system observability decreases as $T$ increases. 
After around $T = 0.1$ ms, convergence becomes unlikely. 
Above this threshold, Newton's method overfits the parameters to the dominant eigenmode, as the dominant mode is effectively overrepresented because the nondominant eigenmodes decay  to negligible levels. 
Consistent with Figure \ref{fig:CondNum}, the inclusion of precursors does not impact the Q-factor for $T<0.1$ ms.
Unfortunately, the condition-limiting effect of the precursors takes effect around when Newton's method begins to already diverge.

\section{Conclusion}
\label{sec:conc}
The feasibility of recovering coupling coefficients $K$ from observational transient data for the multipoint kinetic equations was investigated. 
Kobayashi's MPK formulation was chosen and homogenizing assumptions were made to isolate the effects of $K$. 
The solution to the homogenized MPK equations was analytically derived for two primary cases, one considering no precursors present and one considering a single precursor group.
To enable a simpler, more decoupled form, the solution was restated using the eigenbasis of $K$. 
This form was numerically analyzed using the transient published in \cite{tommasi2019verification}. 
For this transient, it was observed that source sensitivity to the dominant and least dominant eigenmodes at 0.1 ms in a precursor free form of the MPK equations differed by fifteen orders of magnitude, while the introduction of precursors reduced the difference at 0.1 ms to only seven orders of magnitude. 
This indicates that precursors can delay the decay of nondominant eigenmodes, increasing their observability potential in larger observation windows. 

To investigate the feasibility of eigenmode, and therefore coupling coefficient, recovery of $K$, Newton's method was implemented using the loss function defined in this work, for which the gradient and Hessian were analytically derived for the precursor-free and single-precursor cases.
As a demonstration, the eigenvectors were assumed to be known, and only the eigenvalues were to be recovered.  
For the specific transient analyzed, the initial step in the model parameter search generally became less effective as the observation window increased towards $T<0.1$ ms, after which the nondominant eigenmodes became unobservable in comparison to the dominant eigenmode.
The convergence analysis demonstrated that, for this transient, the inclusion of precursors was insufficient to significantly increase the convergence window, as the eigenmode sensitivity magnitudes had already sufficiently diverged prior to the manifestation of delayed neutron effects. 
\hl{Although not investigated here, transients with a smaller deviation from criticality may result in a better conditioned Hessian, thus allowing convergence in larger observation windows. 
Futhermore, thermal systems with larger timescales may also provide a longer observation window for eigenmode recovery.
Alternative transients that more strongly excite nonfundamental spatial modes, such as source-jerk configurations, may provide a more favorable setting for observing the higher order model behavior.} 
\hl{Future work must also include investigation of eigenvector recovery methods.} 
In conclusion, capturing the behavior of nondominant eigenmodes without oversampling the dominant eigenmode is critical to the recoverability of $K$ through transient data.

\section*{Acknowledgment}
The authors acknowledge the MIT Office of Research Computing and Data for providing high performance computing resources that have contributed to the research results reported within this paper.
This work was supported in part by a MathWorks Nuclear Science and Engineering Fellowship at MIT School of Engineering


\bibliographystyle{elsarticle-num} 
\bibliography{refs.bib}

\appendix

\input{appendix}

\end{document}

%% file: appendix.tex
\section{Loss Function Derivation Flowchart}
\label{app:flo-loss}
\begin{figure}[H]
\centering
\includegraphics[width=0.7\textwidth]{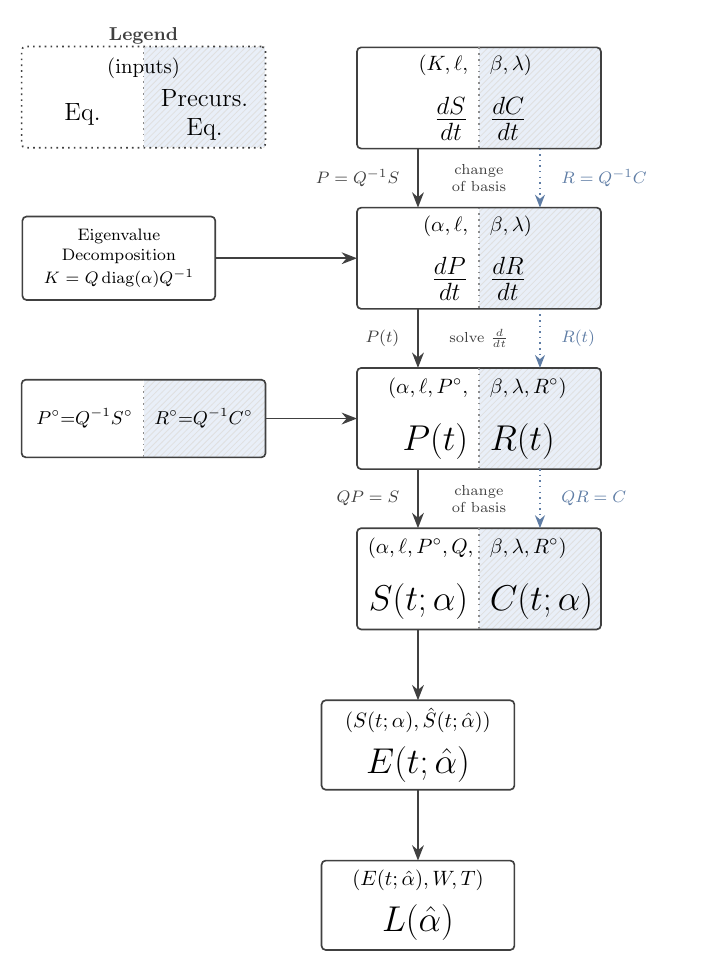}
\caption{\label{fig:lfuncflo}Progression required to derive the loss function.}
\end{figure}

A flowchart showing progression required to derive the loss function, $L(\hat{\alpha})$ is provided in Figure \ref{fig:lfuncflo}.
\section{Loss Function Gradient and Hessian Derivation with no Precursors}
\label{app:noC}
\subsection{Gradient}
\label{app:noC-Grad}
When deriving the gradient, first note that because $W$ is symmetric
\begin{align}
\frac{\partial}{\partial \hat{\alpha}_b} \left(  E(t; \hat{\alpha})^T W E(t; \hat{\alpha})\right) & = \left(\frac{\partial}{\partial \hat{\alpha}_b}E(t; \hat{\alpha})\right)^T  W E(t; \hat{\alpha}) + E(t; \hat{\alpha})^T W \frac{\partial}{\partial \hat{\alpha}_b} E(t; \hat{\alpha}) \\
\frac{\partial}{\partial \hat{\alpha}_b} \left(  E(t; \hat{\alpha})^T W E(t; \hat{\alpha})\right) & = 2 \left(\frac{\partial}{\partial \hat{\alpha}_b}E(t; \hat{\alpha})\right)^T W  E(t; \hat{\alpha}).
\end{align}
Observing that $\frac{\partial}{\partial \hat{\alpha}_b}E(t; \hat{\alpha}) = -\frac{\partial}{\partial \hat{\alpha}_b}\hat{S}(t; \hat{\alpha})$ leads to
\begin{gather}
\frac{\partial}{\partial \hat{\alpha}_b} \left( E(t; \hat{\alpha})^T W E(t; \hat{\alpha})\right)
=
-2 \left( \frac{\partial }{\partial \hat{\alpha}_b}\hat{S}(t; \hat{\alpha}) \right)^T W E(t; \hat{\alpha})
\end{gather}
Also note that
\begin{gather}
\frac{\partial }{\partial \hat{\alpha}_b}\hat{S}(t; \hat{\alpha}) = \frac{t}{\ell}P_b^\circ e^{\frac{\hat{\alpha}_b - 1}{\ell}t} q_b.
\label{Eq:d-anahat}
\end{gather}
Therefore, taking the partial derivative of Equation \ref{Eq:Loss} with respect to eigenvalue $\hat{\alpha}_b$,
\begin{gather}
\frac{\partial}{\partial \hat{\alpha}_b}  L(\hat{\alpha})
=
-2 \int_0^T
\left(\frac{\partial }{\partial \hat{\alpha}_b}\hat{S}(t; \hat{\alpha})\right)^T
W E(t; \hat{\alpha})\,dt.
\label{eq:dLoss}
\end{gather}
Propogating Equation \ref{Eq:d-anahat} into Equation \ref{eq:dLoss},
\begin{align}
\frac{\partial}{\partial \hat{\alpha}_b}  L(\hat{\alpha})
 & =
-2 \int_0^T
\frac{t}{\ell}P_b^\circ e^{\frac{\hat{\alpha}_b - 1}{\ell}t} q_b^T
W \sum_{j=1}^N P_j^\circ \left(e^{\frac{\alpha_j - 1}{\ell}t} -e^{\frac{\hat{\alpha}_j - 1}{\ell}t} \right)q_jdt.
\end{align}
Then, 
\begin{align}
\frac{\partial}{\partial \hat{\alpha}_b}  L(\hat{\alpha})
  =
-\frac{2}{\ell}\sum_{j=1}^N M_{b,j}^\circ
\int_0^T
t
\left(
e^{\frac{\hat{\alpha}_b+\alpha_j-2}{\ell}t}
-
e^{\frac{\hat{\alpha}_b+\hat{\alpha}_j-2}{\ell}t}
\right)dt.
\label{Eq:GradnoI}
\end{align}
where
\begin{gather}
M_{b,j}^\circ \equiv P_b^\circ P_j^\circ q_b^T W q_j,
\end{gather}
which captures the overlap of the eigenstates at the initiation of the transient.
For a simplified final expression, define
\begin{gather}
I_n\left(\gamma, T\right) \equiv \int_0^T t^n e^{\gamma t} dt.
\label{Eq:In}
\end{gather}
Integrals of this type have analytical expressions provided in Section 2.32 of \cite{gradshteyn2014table}.
Then, 
\begin{align}
\frac{\partial}{\partial \hat{\alpha}_b}  L(\hat{\alpha})
  =
-\frac{2}{\ell}\sum_{j=1}^N M_{b,j}^\circ
\left(I_1\left(\frac{\hat{\alpha}_b + \alpha_j - 2}{\ell}, T \right) - I_1\left(\frac{\hat{\alpha}_b + \hat{\alpha}_j - 2}{\ell}, T \right)\right).
\end{align}

\subsection{Hessian}
\label{app:noC-Hess}
To derive the Hessian, differentiate Equation \ref{eq:dLoss} once more with respect to $\hat{\alpha}_c$ to obtain
\begin{align}
\frac{\partial^2}{\partial \hat{\alpha}_c \partial \hat{\alpha}_b}  L(\hat{\alpha})
&=
-2 \int_0^T
\frac{\partial}{\partial \hat{\alpha}_c}
\left[
\left(\frac{\partial }{\partial \hat{\alpha}_b}\hat{S}(t; \hat{\alpha})\right)^T
W E(t; \hat{\alpha})
\right]dt \\
&= -2 \int_0^T
\left(\frac{\partial^2 }{\partial \hat{\alpha}_c \partial \hat{\alpha}_b}\hat{S}(t; \hat{\alpha})\right)^T
W E(t; \hat{\alpha})
-
\left(\frac{\partial }{\partial \hat{\alpha}_b}\hat{S}(t; \hat{\alpha})\right)^T
W
\left(\frac{\partial }{\partial \hat{\alpha}_c}\hat{S}(t; \hat{\alpha})\right)
\,dt.
\label{eq:linalgH}
\end{align}
From Equation \ref{Eq:d-anahat}, differentiating once more gives
\begin{gather}
\frac{\partial^2 }{\partial \hat{\alpha}_c \partial \hat{\alpha}_b}\hat{S}(t; \hat{\alpha})
=
\delta_{b,c}\frac{t^2}{\ell^2}P_b^\circ e^{\frac{\hat{\alpha}_b - 1}{\ell}t} q_b
\end{gather}
where $\delta_{b,c}$ is the Kronecker delta. 
Now, make substitutions into Equation \ref{eq:linalgH} and simplify similar to Equation \ref{Eq:GradnoI} to obtain
\begin{align}
\frac{\partial^2}{\partial \hat{\alpha}_c \partial \hat{\alpha}_b}  L(\hat{\alpha})
&=
-\frac{2}{\ell^2}\delta_{b,c}\sum_{j=1}^N M_{b,j}^\circ
\int_0^T
t^2
\left(
e^{\frac{\hat{\alpha}_b+\alpha_j-2}{\ell}t}
-
e^{\frac{\hat{\alpha}_b+\hat{\alpha}_j-2}{\ell}t}
\right)\,dt 
+\frac{2}{\ell^2}M_{b,c}^\circ
\int_0^T
t^2 e^{\frac{\hat{\alpha}_b+\hat{\alpha}_c-2}{\ell}t}\,dt.
\end{align}
Implementing Equation \ref{Eq:In},
\begin{gather}
\frac{\partial^2}{\partial \hat{\alpha}_c \partial \hat{\alpha}_b}  L(\hat{\alpha})
=
-\frac{2}{\ell^2}\delta_{b,c}\sum_{j=1}^N M_{b,j}^\circ
\left(
I_2\left(\frac{\hat{\alpha}_b+\alpha_j-2}{\ell}, T\right)
-
I_2\left(\frac{\hat{\alpha}_b+\hat{\alpha}_j-2}{\ell}, T\right)
\right)
\label{Eq:HessNoC}\\
+
\frac{2}{\ell^2}M_{b,c}^\circ
I_2\left(\frac{\hat{\alpha}_b+\hat{\alpha}_c-2}{\ell}, T\right).
\notag
\end{gather}

\section{One Precursor Group Derivations}
\label{app:1C}
\subsection{Gradient}
\label{app:1C-grad}
To begin, know that the gradient of the loss function in the case of one precursor has the same structure, such that 
\begin{gather}
\frac{\partial}{\partial \hat{\alpha}_b}  L(\hat{\alpha})
=
-2 \int_0^T
\left(\frac{\partial }{\partial \hat{\alpha}_b}\hat{S}(t; \hat{\alpha})\right)^T
W E(t; \hat{\alpha})\,dt.
\end{gather}
still holds. 
For readability, say that 
\begin{align}
\hat{S}(t;\hat{\alpha}) &=
\sum^N_{j=1}\left(\frac{\left(\hat{\omega}_j^+ + \lambda\right)
\left(\beta P_j^\circ-\left(\hat{\omega}_j^-+ \lambda\right)R_j^\circ\right)} {\beta\left(\hat{\omega}_j^+-\hat{\omega}_j^-\right)} \exp\left(\hat{\omega}_j^+t\right)
 - \frac{\left(\hat{\omega}_j^- + \lambda\right) \left(\beta P_j^\circ-\left(\hat{\omega}_j^++\lambda\right)R_j^\circ\right)}
{\beta\left(\hat{\omega}_j^+-\hat{\omega}_j^-\right)} \exp\left(\hat{\omega}_j^-t\right)
\right)q_j \\
&= \sum^N_{j=1}
\left(
\hat{\xi}_j^+\exp\left(\hat{\omega}_j^+t\right)
-\hat{\xi}_j^-\exp\left(\hat{\omega}_j^-t\right)
\right)q_j .
\end{align}
where
\begin{align}
\hat{\xi}_j^\pm &= \frac{\left(\hat{\omega}_j^\pm+\lambda\right)\left(\beta P_j^\circ-\left(\hat{\omega}_j^\mp+\lambda\right)R_j^\circ\right)}{\beta\left(\hat{\omega}_j^+-\hat{\omega}_j^-\right)}.
\end{align}
Further define
\begin{align}
\hat{\xi}_j^+ &= \frac{\left(\hat{\omega}_j^++\lambda\right)\left(\beta P_j^\circ-\left(\hat{\omega}_j^-+\lambda\right)R_j^\circ\right)}{\beta\left(\hat{\omega}_j^+-\hat{\omega}_j^-\right)}=\frac{\epsilon_j^+\nu_j^-}{\eta_j}=\frac{F_j^+}{\eta_j},\\
\hat{\xi}_j^- &= \frac{\left(\hat{\omega}_j^-+\lambda\right)\left(\beta P_j^\circ-\left(\hat{\omega}_j^++\lambda\right)R_j^\circ\right)}{\beta\left(\hat{\omega}_j^+-\hat{\omega}_j^-\right)}=\frac{\epsilon_j^-\nu_j^+}{\eta_j}=\frac{F_j^-}{\eta_j}.
\end{align}
and
\begin{align}
F_j^\pm &= \epsilon_j^\pm\nu_j^\mp,\\
\epsilon_j^+ &= \hat{\omega}_j^++\lambda,\\
\epsilon_j^- &= \hat{\omega}_j^-+\lambda,\\
\nu_j^- &= \beta P_j^\circ-\left(\hat{\omega}_j^-+\lambda\right)R_j^\circ,\\
\nu_j^+ &= \beta P_j^\circ-\left(\hat{\omega}_j^++\lambda\right)R_j^\circ,\\
\eta_j &= \beta\left(\hat{\omega}_j^+-\hat{\omega}_j^-\right).
\end{align}
Taking the first partial derivative with respect to an estimated eigenvalue $\hat{\alpha}_b$,
\begin{align}
\ddk{}{\hat{\alpha}_b}\hat{S}(t;\hat{\alpha}) = & 
\Bigg( \ddk{\hat{\xi}_b^+}{\hat{\alpha}_b} \exp\left( \hat{\omega}_b^+ t\right) 
+ \hat{\xi}_b^+ \exp\left( \hat{\omega}_b^+ t\right)t \ddk{\hat{\omega}_b^+}{\hat{\alpha}_b}
-  \ddk{\hat{\xi}_b^-}{\hat{\alpha}_b}\exp\left( \hat{\omega}_b^- t\right) 
- \hat{\xi}_b^- \exp\left( \hat{\omega}_b^- t\right) t\ddk{\hat{\omega}_b^-}{\hat{\alpha}_b} \Bigg) q_b \\
&=
\left( 
    \Big( \frac{\eta_b \ddk{F_b^+}{\hat{\alpha}_b}-\ddk{\eta_b}{\hat{\alpha}_b}F_b^+}{\eta_b^2}
    + \frac{F_b^+}{\eta_b}t \ddk{\hat{\omega}_b^+}{\hat{\alpha}_b} \Big) \exp( \hat{\omega}_b^+ t) 
    - \Big( \frac{\eta_b \ddk{F_b^-}{\hat{\alpha}_b} -\ddk{\eta_b}{\hat{\alpha}_b}F_b^-}{\eta_b^2} + \frac{F_b^-}{\eta_b}  t \ddk{\hat{\omega}_b^-}{\hat{\alpha}_b}  \Big) \exp( \hat{\omega}_b^- t)
\right) q_b  \\
&=
\left( \frac{\eta_b \big(\epsilon_b^+\ddk{\nu_b^-}{\hat{\alpha}_b}+ \ddk{\epsilon_b^+}{\hat{\alpha}_b} \nu_b^-\big)-\ddk{\eta_b}{\hat{\alpha}_b}\epsilon_b^+ \nu_b^-}{\eta_b^2} \exp\left( \hat{\omega}_b^+ t\right) 
+ \frac{\epsilon_b^+ \nu_b^-}{\eta_b} \exp\left( \hat{\omega}_b^+ t\right)t \ddk{\hat{\omega}_b^+}{\hat{\alpha}_b}
\right.\\&\left. 
-  \frac{\eta_b \big(\epsilon_b^-\ddk{\nu_b^+}{\hat{\alpha}_b}+ \ddk{\epsilon_b^-}{\hat{\alpha}_b} \nu_b^+\big) -\ddk{\eta_b}{\hat{\alpha}_b}\epsilon_b^- \nu_b^+}{\eta_b^2} \exp\left( \hat{\omega}_b^- t\right) 
- \frac{\epsilon_b^- \nu_b^+}{\eta_b} \exp\left( \hat{\omega}_b^- t\right) t \ddk{\hat{\omega}_b^-}{\hat{\alpha}_b}
\right) q_b ,
\end{align}
as
\begin{gather}
    \ddk{\hat{\xi}_b^+}{\hat{\alpha}_b} 
    = \frac{\eta_b \ddk{F_b^+}{\hat{\alpha}_b}-\ddk{\eta_b}{\hat{\alpha}_b}F_b^+}{\eta_b^2}
    = \frac{\eta_b \big(\epsilon_b^+\ddk{\nu_b^-}{\hat{\alpha}_b}+ \ddk{\epsilon_b^+}{\hat{\alpha}_b} \nu_b^-\big)-\ddk{\eta_b}{\hat{\alpha}_b}\epsilon_b^+ \nu_b^-}{\eta_b^2} \\
    \ddk{\hat{\xi}_b^-}{\hat{\alpha}_b} 
    = \frac{\eta_b \ddk{F_b^-}{\hat{\alpha}_b} -\ddk{\eta_b}{\hat{\alpha}_b}F_b^-}{\eta_b^2}
    = \frac{\eta_b \big(\epsilon_b^-\ddk{\nu_b^+}{\hat{\alpha}_b}+ \ddk{\epsilon_b^-}{\hat{\alpha}_b} \nu_b^+\big) -\ddk{\eta_b}{\hat{\alpha}_b}\epsilon_b^- \nu_b^+}{\eta_b^2} \\
    \ddk{F_b^+}{\hat{\alpha}_b}  
    = \epsilon_b^+\ddk{\nu_b^-}{\hat{\alpha}_b}+ \ddk{\epsilon_b^+}{\hat{\alpha}_b} \nu_b^- \\
    \ddk{F_b^-}{\hat{\alpha}_b} = \epsilon_b^-\ddk{\nu_b^+}{\hat{\alpha}_b}+ \ddk{\epsilon_b^-}{\hat{\alpha}_b} \nu_b^+ 
\end{gather}
and
\begin{align}
    \ddk{\epsilon_b^\pm}{\hat{\alpha}_b}  &= \ddk{\epsilon_b^\pm}{\hat{\omega}_b^\pm}\ddk{\hat{\omega}_b^\pm}{\hat{\alpha}_b} = 1\ddk{\hat{\omega}_b^\pm}{\hat{\alpha}_b} \\
    \ddk{\nu_b^\pm}{\hat{\alpha}_b} &= \ddk{\nu_b^\pm}{\hat{\omega}_b^\pm}\ddk{\hat{\omega}_b^\pm}{\hat{\alpha}_b} = -R_b^\circ \ddk{\hat{\omega}_b^\pm}{\hat{\alpha}_b}\\
    \ddk{\eta_b}{\hat{\alpha}_b} &=  \beta\left(\ddk{\hat{\omega}_b^+}{\hat{\alpha}_b} - \ddk{\hat{\omega}_b^-}{\hat{\alpha}_b}\right) .
\end{align}
and
\begin{align}
    \ddk{\hat{\omega}_b^\pm}{\hat{\alpha}_b} = 
    \frac{1}{2}
    \left(
    \frac{1-\beta}{\ell}
    \pm
    \frac{1}{2}\left[
    \left(\lambda-\frac{(1-\beta)\hat{\alpha}_b-1}{\ell}\right)^2
    +\frac{4\lambda(\hat{\alpha}_b-1)}{\ell}
    \right]^{-1/2}
    \left[2\left(\lambda-\frac{(1-\beta)\hat{\alpha}_b-1}{\ell}\right) \left(-\frac{1-\beta}{\ell}\right)
    +\frac{4\lambda}{\ell}
    \right]
    \right) \\
    = 
    \frac{1}{2}\left(
    \frac{1-\beta}{\ell}
    \pm
    \dfrac{
    \left[2\left(\lambda-\frac{(1-\beta)\hat{\alpha}_b-1}{\ell}\right) \left(-\frac{1-\beta}{\ell}\right)
    +\frac{4\lambda}{\ell}
    \right]}{
    2\left[
    \left(\lambda-\frac{(1-\beta)\hat{\alpha}_b-1}{\ell}\right)^2
    +\frac{4\lambda(\hat{\alpha}_b-1)}{\ell}
    \right]^{1/2}
    } \right)
\end{align}

\subsection{Hessian}
\label{app:1C-Hess}
To begin, know that the Hessian of the loss function in the case of one precursor has the same structure, such that 
\begin{align}
\frac{\partial^2}{\partial \hat{\alpha}_c \partial \hat{\alpha}_b}  L(\hat{\alpha})
&= -2 \int_0^T \Bigg(
\Big(\frac{\partial^2 }{\partial \hat{\alpha}_c \partial \hat{\alpha}_b}\hat{S}(t; \hat{\alpha})\Big)^T
W E(t; \hat{\alpha})
-
\Big(\frac{\partial }{\partial \hat{\alpha}_b}\hat{S}(t; \hat{\alpha})\Big)^T
W
\Big(\frac{\partial }{\partial \hat{\alpha}_c}\hat{S}(t; \hat{\alpha})\Big) 
\Bigg) \,dt,
\label{eq:1CHess}
\end{align}
still holds. 
Taking the partial derivative of the source once more with respect to an eigenvalue, as needed in \ref{eq:1CHess},
\begin{align} 
\dkdk{}{\hat{\alpha}_c}{\hat{\alpha}_b}\hat{S}(t;\hat{\alpha})
=& \ddk{}{\hat{\alpha}_c}
\left[ \ddk{\hat{\xi}_b^+}{\hat{\alpha}_b} \exp\left( \hat{\omega}_b^+ t\right) 
+ \hat{\xi}_b^+ \exp\left( \hat{\omega}_b^+ t\right)t \ddk{\hat{\omega}_b^+}{\hat{\alpha}_b}
-  \ddk{\hat{\xi}_b^-}{\hat{\alpha}_b}\exp\left( \hat{\omega}_b^- t\right) 
- \hat{\xi}_b^- \exp\left( \hat{\omega}_b^- t\right) t\ddk{\hat{\omega}_b^-}{\hat{\alpha}_b}
\right] q_b 
\end{align}
Recall that $\hat{\omega}_j^\pm$ is a function of no other eigenvalues except $\hat{\alpha}_j$. 
Therefore, when $b\neq c$, $\dkdk{}{\hat{\alpha}_c}{\hat{\alpha}_b}\hat{S}(t;\hat{\alpha}) = 0$, thus all mixed derivatives of $\hat{S}(t;\hat{\alpha})$ are zero. 
Only pure derivatives are nonzero. 
For ease of notation, only use subscript $b$ in the following derivation. 
Define $G_b^\pm$ to be 
\begin{align}
    G_b^\pm = \ddk{ \hat{\xi}_b^\pm}{\hat{\alpha}_b} \exp\left( \hat{\omega}_b^\pm t\right) 
    +  \hat{\xi}_b^\pm \exp\left( \hat{\omega}_b^\pm t\right)t \ddk{\hat{\omega}_b^\pm}{\hat{\alpha}_b}
\end{align}
such that
\begin{align} 
\dkdk{}{\hat{\alpha}_c}{\hat{\alpha}_b}\hat{S}(t;\hat{\alpha})
=& \delta_{b,c} \ddk{}{\hat{\alpha}_c}
\left[ \ddk{\hat{\xi}_b^+}{\hat{\alpha}_b} \exp\left( \hat{\omega}_b^+ t\right) 
+ \hat{\xi}_b^+ \exp\left( \hat{\omega}_b^+ t\right)t \ddk{\hat{\omega}_b^+}{\hat{\alpha}_b}
-  \ddk{\hat{\xi}_b^-}{\hat{\alpha}_b}\exp\left( \hat{\omega}_b^- t\right) 
- \hat{\xi}_b^- \exp\left( \hat{\omega}_b^- t\right) t\ddk{\hat{\omega}_b^-}{\hat{\alpha}_b}
\right] q_b \\
=& \delta_{b,c}\ddk{}{\hat{\alpha}_b} \left[G_b^+ - G_b^-\right]q_b. \\
\end{align}
Taking the first partial derivative,
\begin{align}
    \ddk{G_b^\pm}{\hat{\alpha}_b} = &\,\dsqdk{ \hat{\xi}_b^\pm}{\hat{\alpha}_b} \exp( \hat{\omega}_b^\pm t) 
    + \ddk{ \hat{\xi}_b^\pm}{\hat{\alpha}_b} \exp( \hat{\omega}_b^\pm t) t \ddk{\hat{\omega}_b^\pm}{\hat{\alpha}_b} \\
    &+ \ddk{ \hat{\xi}_b^\pm}{\hat{\alpha}_b} \exp( \hat{\omega}_b^\pm t)t \ddk{\hat{\omega}_b^\pm}{\hat{\alpha}_b}
    +  \hat{\xi}_b^\pm \exp( \hat{\omega}_b^\pm t)\Big(t \ddk{\hat{\omega}_b^\pm}{\hat{\alpha}_b}\Big)^2
    +  \hat{\xi}_b^\pm \exp( \hat{\omega}_b^\pm t)t \dsqdk{\hat{\omega}_b^\pm}{\hat{\alpha}_b} \\
    = &\,\dsqdk{ \hat{\xi}_b^\pm}{\hat{\alpha}_b} \exp( \hat{\omega}_b^\pm t) 
    + 2\ddk{ \hat{\xi}_b^\pm}{\hat{\alpha}_b} \exp( \hat{\omega}_b^\pm t) t \ddk{\hat{\omega}_b^\pm}{\hat{\alpha}_b} 
    +  \hat{\xi}_b^\pm \exp( \hat{\omega}_b^\pm t)\Big(t \ddk{\hat{\omega}_b^\pm}{\hat{\alpha}_b}\Big)^2
    +  \hat{\xi}_b^\pm \exp( \hat{\omega}_b^\pm t)t \dsqdk{\hat{\omega}_b^\pm}{\hat{\alpha}_b}
\end{align}
where
\begin{align}
    \dsqdk{\hat{\xi}_b^+}{\hat{\alpha}_b}  &= 
    \frac{1}{\eta_b^4}
    \Bigg( \eta_b^2\ddk{}{\hat{\alpha}_b} \Big[
    \eta_b \ddk{F_b^+}{\hat{\alpha}_b}-\ddk{\eta_b}{\hat{\alpha}_b}F_b^+\Big] 
    - 2\eta_b\ddk{\eta_b}{\hat{\alpha}_b}\Big(
    \eta_b \ddk{F_b^+}{\hat{\alpha}_b}-\ddk{\eta_b}{\hat{\alpha}_b}F_b^+\Big) \Bigg) \\
    &= \frac{1}{\eta_b^4}
    \Bigg( \eta_b^2\Big(
    \ddk{\eta_b}{\hat{\alpha}_b} \ddk{F_b^+}{\hat{\alpha}_b}
    + \eta_b \dsqdk{F_b^+}{\hat{\alpha}_b}
    -\dsqdk{\eta_b}{\hat{\alpha}_b}F_b^+
    -\ddk{\eta_b}{\hat{\alpha}_b}\ddk{\epsilon_b^+}{\hat{\alpha}_b} \nu_b^-
    -\ddk{\eta_b}{\hat{\alpha}_b}\epsilon_b^+ \ddk{\nu_b^-}{\hat{\alpha}_b} \Big)
    - 2\eta_b\ddk{\eta_b}{\hat{\alpha}_b}\Big(
    \eta_b \ddk{F_b^+}{\hat{\alpha}_b}-\ddk{\eta_b}{\hat{\alpha}_b}F_b^+\Big) \Bigg)\\
    &= \frac{1}{\eta_b^4}
    \Bigg( \eta_b^2\Big(-
    \ddk{\eta_b}{\hat{\alpha}_b} \ddk{F_b^+}{\hat{\alpha}_b}
    + \eta_b \dsqdk{F_b^+}{\hat{\alpha}_b}
    -\dsqdk{\eta_b}{\hat{\alpha}_b}F_b^+
    -\ddk{\eta_b}{\hat{\alpha}_b}\ddk{F_b^+}{\hat{\alpha}_b} \Big)
    + 2\eta_b\Big(\ddk{\eta_b}{\hat{\alpha}_b}\Big)^2F_b^+ \Bigg)\\
    &= \frac{1}{\eta_b^4}
    \Bigg( \eta_b^2\Big(-2
    \ddk{\eta_b}{\hat{\alpha}_b} \ddk{F_b^+}{\hat{\alpha}_b}
    + \eta_b \dsqdk{F_b^+}{\hat{\alpha}_b}
    -\dsqdk{\eta_b}{\hat{\alpha}_b}F_b^+\Big) 
    + 2\eta_b\Big(\ddk{\eta_b}{\hat{\alpha}_b}\Big)^2F_b^+\Bigg)\\
    &= \frac{1}{\eta_b} \dsqdk{F_b^+}{\hat{\alpha}_b} 
    -\frac{2}{\eta_b^2}\ddk{\eta_b}{\hat{\alpha}_b} \ddk{F_b^+}{\hat{\alpha}_b}
    -\frac{1}{\eta_b^2}\dsqdk{\eta_b}{\hat{\alpha}_b}F_b^+  
    + \frac{2}{\eta_b^3}\Big(\ddk{\eta_b}{\hat{\alpha}_b}\Big)^2F_b^+ \\
    \dsqdk{\hat{\xi}_b^-}{\hat{\alpha}_b}  &= 
    \frac{1}{\eta_b^4}
    \Bigg( \eta_b^2\ddk{}{\hat{\alpha}_b} \Big[
    \eta_b \ddk{F_b^-}{\hat{\alpha}_b}-\ddk{\eta_b}{\hat{\alpha}_b}F_b^-\Big] - 2\eta_b\ddk{\eta_b}{\hat{\alpha}_b}\Big(
    \eta_b \ddk{F_b^-}{\hat{\alpha}_b}-\ddk{\eta_b}{\hat{\alpha}_b}F_b^-\Big) \Bigg) \\
    &= \frac{1}{\eta_b} \dsqdk{F_b^-}{\hat{\alpha}_b} 
    -\frac{2}{\eta_b^2}\ddk{\eta_b}{\hat{\alpha}_b} \ddk{F_b^-}{\hat{\alpha}_b}
    -\frac{1}{\eta_b^2}\dsqdk{\eta_b}{\hat{\alpha}_b}F_b^-  
    + \frac{2}{\eta_b^3}\Big(\ddk{\eta_b}{\hat{\alpha}_b}\Big)^2F_b^- \\
    \dsqdk{F_b^\pm}{\hat{\alpha}_b} &= 
    \epsilon_b^\pm\dsqdk{\nu_b^\mp}{\hat{\alpha}_b}
    + 2\ddk{\epsilon_b^\pm}{\hat{\alpha}_b}\ddk{\nu_b^\mp}{\hat{\alpha}_b}
    + \dsqdk{\epsilon_b^\pm}{\hat{\alpha}_b} \nu_b^\mp \\
\end{align}
Propagating the second partial derivative, 
\begin{align}
    \dsqdk{\epsilon_b^\pm}{\hat{\alpha}_b}  &= \dsqdk{\hat{\omega}_b^\pm}{\hat{\alpha}_b} \\
    \dsqdk{\nu_b^\pm}{\hat{\alpha}_b} &= -R_b^\circ \dsqdk{\hat{\omega}_b^\pm}{\hat{\alpha}_b}\\
    \dsqdk{\eta_b}{\hat{\alpha}_b}  &= \beta\left(\dsqdk{\hat{\omega}_b^+}{\hat{\alpha}_b} - \dsqdk{\hat{\omega}_b^-}{\hat{\alpha}_b}\right) .
\end{align} 

Finally, 
\begin{align}
    \dkdk{\hat{\omega}_b^\pm}{\hat{\alpha}_b}{\hat{\alpha}_c} &= 0\\
    \dsqdk{\hat{\omega}_b^\pm} {\hat{\alpha}_b} &= 
    \pm
    \frac{1}{4}
    \ddk{}{\hat{\alpha}_b}
    \left[
    \left(2\left(\lambda-\frac{(1-\beta)\hat{\alpha}_b-1}{\ell}\right) \left(-\frac{1-\beta}{\ell}\right)
    +\frac{4\lambda}{\ell}
    \right) 
    \left(
    \left(\lambda-\frac{(1-\beta)\hat{\alpha}_b-1}{\ell}\right)^2
    +\frac{4\lambda(\hat{\alpha}_b-1)}{\ell}
    \right)^{-1/2}
    \right] \\
    =& \pm\frac{1}{4}\Bigg( \ddk{U}{\hat{\alpha}_b}V + U\ddk{V}{\hat{\alpha}_b}\Bigg)
\end{align}
where
\begin{align}
    U &= \Bigg(
    \Big(\lambda-\frac{(1-\beta)\hat{\alpha}_b-1}{\ell}\Big)^2
    +\frac{4\lambda(\hat{\alpha}_b-1)}{\ell}
    \Bigg)^{-1/2}\\
    V &= 2\left(\lambda-\frac{(1-\beta)\hat{\alpha}_b-1}{\ell}\right) \left(-\frac{1-\beta}{\ell}\right)
    +\frac{4\lambda}{\ell}\\
    \ddk{U}{\hat{\alpha}_b} &= \frac{-1}{2}\Bigg(
    \Big(\lambda-\frac{(1-\beta)\hat{\alpha}_b-1}{\ell}\Big)^2
    +\frac{4\lambda(\hat{\alpha}_b-1)}{\ell}
    \Bigg)^{-3/2}  V\\
    \ddk{V}{\hat{\alpha}_b} &= 2 \left(-\frac{1-\beta}{\ell}\right)^2
\end{align}

A collection of key definitions and final forms required for the Hessian calculation can be found below.
\begin{align}
    U &= \Bigg(
    \Big(\lambda-\frac{(1-\beta)\hat{\alpha}_b-1}{\ell}\Big)^2
    +\frac{4\lambda(\hat{\alpha}_b-1)}{\ell}
    \Bigg)^{-1/2}\\
    V &= 2\left(\lambda-\frac{(1-\beta)\hat{\alpha}_b-1}{\ell}\right) \left(-\frac{1-\beta}{\ell}\right)
    +\frac{4\lambda}{\ell}\\
    \ddk{U}{\hat{\alpha}_b} &= \frac{-1}{2}\Bigg(
    \Big(\lambda-\frac{(1-\beta)\hat{\alpha}_b-1}{\ell}\Big)^2
    +\frac{4\lambda(\hat{\alpha}_b-1)}{\ell}
    \Bigg)^{-3/2}  V\\
    \ddk{V}{\hat{\alpha}_b} &= 2 \left(-\frac{1-\beta}{\ell}\right)^2
\end{align}
\begin{align}
    \dkdk{\hat{\omega}_b^\pm}{\hat{\alpha}_b}{\hat{\alpha}_c} &= 0\\
    \dsqdk{\hat{\omega}_b^\pm} {\hat{\alpha}_b} 
    =& \pm\frac{1}{4}\Bigg( \ddk{U}{\hat{\alpha}_b}V + U\ddk{V}{\hat{\alpha}_b}\Bigg)
\end{align} 
\begin{align}
    \dsqdk{\epsilon_b^\pm}{\hat{\alpha}_b}  &= \dsqdk{\hat{\omega}_b^\pm}{\hat{\alpha}_b} \\
    \dsqdk{\nu_b^\pm}{\hat{\alpha}_b} &= -R_b^\circ \dsqdk{\hat{\omega}_b^\pm}{\hat{\alpha}_b}\\
    \dsqdk{\eta_b}{\hat{\alpha}_b}  &= \beta\left(\dsqdk{\hat{\omega}_b^+}{\hat{\alpha}_b} - \dsqdk{\hat{\omega}_b^-}{\hat{\alpha}_b}\right) .
\end{align}
 \begin{align}
    F_j^\pm &= \epsilon_j^\pm\nu_j^\mp \\
    \ddk{F_b^+}{\hat{\alpha}_b}  
    &= \epsilon_b^+\ddk{\nu_b^-}{\hat{\alpha}_b}+ \ddk{\epsilon_b^+}{\hat{\alpha}_b} \nu_b^- \\
    \ddk{F_b^-}{\hat{\alpha}_b} &= \epsilon_b^-\ddk{\nu_b^+}{\hat{\alpha}_b}+ \ddk{\epsilon_b^-}{\hat{\alpha}_b} \nu_b^+ \\
    \dsqdk{F_b^\pm}{\hat{\alpha}_b} &= 
    \epsilon_b^\pm\dsqdk{\nu_b^\mp}{\hat{\alpha}_b}
    + 2\ddk{\epsilon_b^\pm}{\hat{\alpha}_b}\ddk{\nu_b^\mp}{\hat{\alpha}_b}
    + \dsqdk{\epsilon_b^\pm}{\hat{\alpha}_b} \nu_b^\mp
\end{align}
\begin{align}
    \dsqdk{\hat{\xi}_b^+}{\hat{\alpha}_b} &= \frac{1}{\eta_b} \dsqdk{F_b^+}{\hat{\alpha}_b} 
    -\frac{2}{\eta_b^2}\ddk{\eta_b}{\hat{\alpha}_b} \ddk{F_b^+}{\hat{\alpha}_b}
    -\frac{1}{\eta_b^2}\dsqdk{\eta_b}{\hat{\alpha}_b}F_b^+  
    + \frac{2}{\eta_b^3}\Big(\ddk{\eta_b}{\hat{\alpha}_b}\Big)^2F_b^+ \\
    \dsqdk{\hat{\xi}_b^-}{\hat{\alpha}_b}  &= 
    \frac{1}{\eta_b} \dsqdk{F_b^-}{\hat{\alpha}_b} 
    -\frac{2}{\eta_b^2}\ddk{\eta_b}{\hat{\alpha}_b} \ddk{F_b^-}{\hat{\alpha}_b}
    -\frac{1}{\eta_b^2}\dsqdk{\eta_b}{\hat{\alpha}_b}F_b^-  
    + \frac{2}{\eta_b^3}\Big(\ddk{\eta_b}{\hat{\alpha}_b}\Big)^2F_b^-
\end{align}

\begin{align}
    G_b^\pm =&\, \ddk{ \hat{\xi}_b^\pm}{\hat{\alpha}_b} \exp\left( \hat{\omega}_b^\pm t\right) 
    +  \hat{\xi}_b^\pm \exp\left( \hat{\omega}_b^\pm t\right)t \ddk{\hat{\omega}_b^\pm}{\hat{\alpha}_b} \\
    \ddk{G_b^\pm}{\hat{\alpha}_b} = &\,\dsqdk{ \hat{\xi}_b^\pm}{\hat{\alpha}_b} \exp( \hat{\omega}_b^\pm t) 
    + 2\ddk{ \hat{\xi}_b^\pm}{\hat{\alpha}_b} \exp( \hat{\omega}_b^\pm t) t \ddk{\hat{\omega}_b^\pm}{\hat{\alpha}_b} 
    +  \hat{\xi}_b^\pm \exp( \hat{\omega}_b^\pm t)\Big(t \ddk{\hat{\omega}_b^\pm}{\hat{\alpha}_b}\Big)^2
    +  \hat{\xi}_b^\pm \exp( \hat{\omega}_b^\pm t)t \dsqdk{\hat{\omega}_b^\pm}{\hat{\alpha}_b}
\end{align}
\begin{align} 
\dkdk{}{\hat{\alpha}_c}{\hat{\alpha}_b}\hat{S}(t;\hat{\alpha})
=& \delta_{b,c}\ddk{}{\hat{\alpha}_b} \left[G_b^+ - G_b^-\right]q_b.
\end{align} 
\begin{align}
\frac{\partial^2}{\partial \hat{\alpha}_c \partial \hat{\alpha}_b}  L(\hat{\alpha})
&= -2 \int_0^T \Bigg(
\Big(\frac{\partial^2 }{\partial \hat{\alpha}_c \partial \hat{\alpha}_b}\hat{S}(t; \hat{\alpha})\Big)^T
W E(t; \hat{\alpha})
-
\Big(\frac{\partial }{\partial \hat{\alpha}_b}\hat{S}(t; \hat{\alpha})\Big)^T
W
\Big(\frac{\partial }{\partial \hat{\alpha}_c}\hat{S}(t; \hat{\alpha})\Big) 
\Bigg) \,dt.
\end{align}

%% file: refs.bib
@book{wakabayashi2023introduction,
  title={Introduction to nuclear reactor experiments},
  author={Wakabayashi, Genichiro and Yamada, Takahiro and Endo, Tomohiro and Pyeon, Cheol Ho},
  year={2023},
  publisher={Springer}
}

@article{monteiro2024differential,
  title={A differential equation for inverse point kinetics},
  author={Monteiro, Vinicius M and Lima, {\'A}lefe Calebe Ribeiro and Martinez, Aquilino S and Gon{\c{c}}alves, Alessandro C},
  journal={Nuclear Engineering and Design},
  volume={421},
  pages={113094},
  year={2024},
  publisher={Elsevier}
}

@techreport{ferney2024benchmarking,
  title={Benchmarking of Different Inverse Point Kinetics Implementations for an Autocorrected Reactimeter Algorithm},
  author={Ferney, Paul Alexandre and DeHart, Mark D},
  year={2024},
  institution={Idaho National Laboratory (INL), Idaho Falls, ID (United States)}
}

@techreport{avery1958theory,
  title={Theory of coupled reactors},
  author={Avery, R},
  year={1958},
  institution={Argonne National Lab., Lemont, Ill.}
}

@article{kobayashi1992rigorous,
  title={Rigorous derivation of multi-point reactor kinetics equations with explicit dependence on perturbation},
  author={Kobayashi, Keisuke},
  journal={Journal of nuclear science and technology},
  volume={29},
  number={2},
  pages={110--120},
  year={1992},
  publisher={Taylor \& Francis}
}

@article{shimjith2010space,
  title={Space--time kinetics modeling of advanced heavy water reactor for control studies},
  author={Shimjith, SR and Tiwari, AP and Naskar, M and Bandyopadhyay, B},
  journal={Annals of nuclear energy},
  volume={37},
  number={3},
  pages={310--324},
  year={2010},
  publisher={Elsevier}
}

@inproceedings{Dulla2017Multipoint,
  author    = {Sandra Dulla and Piero Ravetto and Paolo Saracco},
  title     = {Some New Thoughts on the Multipoint Method for Reactor Physics Applications},
  booktitle = {Proceedings of the International Conference on Mathematics \& Computational Methods Applied to Nuclear Science \& Engineering (M\&C 2017)},
  address   = {Jeju, Korea},
  month     = apr,
  year      = {2017},
  note      = {On USB proceedings},
  institution = {Politecnico di Torino and INFN}
}

@article{Hui03082025,
author = {Jiuwu Hui},
title = {Development of an Extended State Observer for Monitoring of a PWR Based on a Two-Point Kinetic Model},
journal = {Nuclear Technology},
volume = {211},
number = {8},
pages = {1699--1722},
year = {2025},
publisher = {Taylor \& Francis},
doi = {10.1080/00295450.2024.2426410},
}

@article{ansarifar2017adaptive,
  title={Adaptive robust control for axial offset in the PWR nuclear reactors based on the multipoint reactor model during load-following operation},
  author={Ansarifar, GR and others},
  journal={Annals of Nuclear Energy},
  volume={103},
  pages={251--264},
  year={2017},
  publisher={Elsevier}
}

@techreport{potenza1966spert,
  author      = {Potenza, R. M.},
  title       = {Quarterly Technical Report SPERT Project},
  institution = {National Reactor Testing Station, U.S. Atomic Energy Commission},
  number      = {IDO-17206},
  year        = {1966}
}

@article{jarrah2018experimental,
  title={Experimental validation of the neutronic parameters in the Jordan subcritical assembly},
  author={Jarrah, Ibrahim and Malkawi, Salaheddin and Radaideh, Majdi I and Aldebie, Faisal and El-Mustafah, Mohammad},
  journal={Progress in Nuclear Energy},
  volume={108},
  pages={71--80},
  year={2018},
  publisher={Elsevier}
}

@article{tommasi2019verification,
  title={Verification of iterative matrix solutions for multipoint kinetics equations},
  author={Tommasi, Jean and Palmiotti, Giuseppe},
  journal={Annals of Nuclear Energy},
  volume={124},
  pages={357--371},
  year={2019},
  publisher={Elsevier}
}

@article{keepin1965physics,
  title={Physics of nuclear kinetics},
  author={Keepin, G Robert},
  journal={(No Title)},
  year={1965}
}

@article{rudstam2002delayed,
  title={Delayed neutron data for the major actinides},
  author={Rudstam, G and Finck, Ph and Filip, A and D’angelo, A and McKnight, RD},
  journal={A report by the Working Party on International Evaluation Co-operation of the OECD NEA Nuclear Science Committee},
  volume={6},
  year={2002}
}

@book{gradshteyn2014table,
  title={Table of integrals, series, and products},
  author={Gradshteyn, Izrail Solomonovich and Ryzhik, Iosif Moiseevich},
  year={2014},
  publisher={Academic press}
}

@article{duderstadt1975nuclear,
  title={Nuclear reactor analysis},
  author={Duderstadt, James J and Hamilton, Louis J},
  year={1975},
  publisher={John Wiley and Sons, Inc., New York}
}

@book{nesterov2018lectures,
  title={Lectures on convex optimization},
  author={Nesterov, Yurii and others},
  volume={137},
  year={2018},
    publisher={Springer}
}

@book{burden1985numerical,
  title={Numerical analysis},
  author={Burden, Richard L and Faires, Douglas J},
  year={1985},
  publisher={PWS publishing company}
}

@techreport{MUSE_D8_2005,
  title        = {MUSE Final Report D8: The MUSE Experiments for Sub-critical Neutronics Validation},
  institution  = {European Community, 5th Euratom Framework Programme},
  author       = {Mellier, F. and Kloosterman, Jan Leen and Soule, R. and others},
  year         = {2005},
  month        = {June},
  number       = {D8},
  address      = {Cadarache, France},
  note         = {Contract No. FIKW-CT-2000-00063},
}

@phdthesis{Valocchi_2020,
  author       = {Valocchi, Giorgio},
  title        = {Modélisation neutronique de coeurs complexes en cinétique multidimensionnelle : application au coeur ASTRID faible vidange et à la préparation d'expériences de validation},
  school       = {Aix-Marseille Université},
  year         = {2020},
  month        = {May},
  type         = {Thèse de doctorat},
  address      = {Marseille, France},
  doi          = {10.70675/a9ef9361zd53ez466ez96dczf33f2486ef95},
  url          = {https://theses.fr/2020AIXM0101},
  director     = {Tommasi, Jean and Ravetto, Piero}
}

@book{BoydVandenberghe2004,
  author    = {Stephen Boyd and Lieven Vandenberghe},
  title     = {Convex Optimization},
  publisher = {Cambridge University Press},
  year      = {2004},
  isbn      = {9780521833783}
}

@article{ShirYehudayoff2020,
  author  = {Ofer M. Shir and Amir Yehudayoff},
  title   = {On the Covariance-Hessian Relation in Evolution Strategies},
  journal = {Theoretical Computer Science},
  volume  = {801},
  pages   = {157--174},
  year    = {2020},
  doi     = {10.1016/j.tcs.2019.09.002},
  url     = {https://arxiv.org/abs/1806.03674}
}

@article{valocchi2020reduced,
  title={Reduced order models in reactor kinetics: A comparison between point kinetics and multipoint kinetics},
  author={Valocchi, G and Tommasi, J and Ravetto, P},
  journal={Annals of nuclear energy},
  volume={147},
  pages={107702},
  year={2020},
  publisher={Elsevier}
}

@article{HUI2025128006,
title = {Adaptive Kalman filter-based real-time distribution estimation of unmeasured system states for an advanced passive reactor core under optimized mechanical shim control strategy},
journal = {Expert Systems with Applications},
volume = {286},
pages = {128006},
year = {2025},
issn = {0957-4174},
doi = {https://doi.org/10.1016/j.eswa.2025.128006},
author = {Jiuwu Hui},
}
